\documentclass[twocolumn]{aastex62}

\newcommand{\myemail}{speacock@lpl.arizona.edu}

\usepackage{romanbar}
\usepackage{appendix}
\usepackage[version=4]{mhchem}

\accepted{October 17, 2019}

\submitjournal{ApJ}

\shorttitle{EUV-IR Spectrum of GJ 832, GJ 176, GJ 436}
\shortauthors{Peacock et al.}

\begin{document}

\title{PREDICTING THE EXTREME ULTRAVIOLET RADIATION ENVIRONMENT OF EXOPLANETS AROUND LOW-MASS STARS: GJ 832, GJ 176, GJ 436}

\email{\myemail}

\author{Sarah Peacock}
\affil{University of Arizona, Lunar and Planetary Laboratory, 1629 E University Boulevard, Tucson, AZ 85721, USA}

\author{Travis Barman}
\affiliation{University of Arizona, Lunar and Planetary Laboratory, 1629 E University Boulevard, Tucson, AZ 85721, USA}
\author{Evgenya L. Shkolnik}
\affil{School of Earth and Space Exploration, Arizona State University, Tempe, AZ 85281, USA}
\author{Peter H. Hauschildt}
\affil{Hamburger Sternwarte, Gojenbergsweg 112, D-21029 Hamburg, Germany}
\author{E. Baron}
\affil{Homer L. Dodge Department of Physics and Astronomy, University of Oklahoma, 440 W. Brooks, Rm 100, Norman, OK 73019-2061 USA}
\affil{Hamburger Sternwarte, Gojenbergsweg 112, D-21029 Hamburg, Germany}
\author{Birgit Fuhrmeister}
\affil{Hamburger Sternwarte, Gojenbergsweg 112, D-21029 Hamburg, Germany}

\begin{abstract}

Correct estimates of stellar extreme ultraviolet (EUV; 100 -- 1170 \AA) flux are important for studying the photochemistry and stability of exoplanet atmospheres, as EUV radiation ionizes hydrogen and contributes to the heating, expansion, and potential escape of a planet's upper atmosphere. Contamination from interstellar hydrogen makes observing EUV emission from M stars particularly difficult, and impossible past 100 pc, and necessitates other means to predict the flux in this wavelength regime. We present EUV -- infrared (100 \AA \ -- 5.5 $\micron$) synthetic spectra computed with the PHOENIX atmospheric code of three early M dwarf planet hosts: GJ 832 (M1.5 V), GJ 176 (M2.5 V), and GJ 436 (M3.5 V). These one-dimensional semiempirical nonlocal thermodynamic equilibrium models include simple temperature prescriptions for the stellar chromosphere and transition region, from where ultraviolet (UV; 100 -- 3008 \AA) fluxes originate. We guide our models with \textit{Hubble Space Telescope} far- and near-UV spectra and discuss the ability to constrain these models using \textit{Galaxy Evolution Explorer} UV photometry. Our models closely reproduce the observations and predict the unobservable EUV spectrum at a wavelength resolution of $<$0.1 \AA. The temperature profiles that best reproduce the observations for all three stars are described by nearly the same set of parameters, suggesting that early M type stars may have similar thermal structures in their upper atmospheres. With an impending UV observation gap and the scarcity of observed EUV spectra for stars less luminous and more distant than the Sun, upper-atmosphere models such as these are important for providing realistic spectra across short wavelengths and for advancing our understanding of the effects of radiation on planets orbiting M stars.

\end{abstract}


\NewPageAfterKeywords

\section{Introduction}

The majority of known terrestrial-sized exoplanets located within the canonical habitable zone are found orbiting M stars \citep{shields2016}, including the most nearby, Proxima Centauri b \citep{anglada2016} and three of the seven TRAPPIST-1 system planets \citep{gillon2017}. Due to their cool effective temperatures ($\sim$3,000 K), the canonical habitable zone around M stars is close in, extending from 0.1 to 0.4 au. Their low masses and luminosities allow for easier detection of small, transiting rocky planets orbiting at small separations, occurring at a rate of 0.1 to 0.6 planets per star \citep{dressing2013,kopparapu2013,dressing2015}. An increasing number of exoplanet detections around low-mass stars, particularly those located at short periods, motivates the need to understand the properties of low-mass parent stars.

The stellar ultraviolet (UV, 100 -- 3008 \AA) environment around M stars strongly impacts the formation, evolution, and chemistry of close-in exoplanet atmospheres. Due to phenomena occurring in the outer layers of these stars, most M stars have significant long-term UV variability leading to large amounts of short wavelength emission during their active periods \citep{shkolnik2014,loyd2018a,loyd2018b,schneider2018}. The increased levels of UV radiation can alter the chemical composition of planetary atmospheres and lead to significant mass loss, with different wavelengths in a stellar spectrum driving heating and chemistry in different layers of a planet's atmosphere. 

Stellar far-UV (FUV; 1150 -- 1700 \AA) radiation photodissociates molecules including CO$_2$, CH$_4$, and H$_2$O in the upper atmospheres of planets \citep{segura2005, hu2012, moses2014,rugheimer2015,loyd2016}. For example, the strongest emitting line in the FUV, Lyman $\alpha$ (Ly$\alpha$; 1215.7 \AA), will generate hydrocarbon hazes in the upper atmospheric layers as a direct result of dissociating methane in a planet's ionosphere \citep{trainer2006}. At longer wavelengths, stellar near-UV (NUV; 1680 -- 3008 \AA) flux dissociates both O$_{2}$ and O$_{3}$. Analyzing single and repeated UV flare events and from the M dwarf star, AD Leo, \cite{segura2010} and \cite{Tilley2019} found that the combination of UV radiation and protons is capable of depleting almost all of the O$_3$ on an Earth-like planet in an M star's habitable zone in under 10 years. 

Close-in exoplanets become vulnerable to mass loss as stellar extreme-UV (EUV; 100 -- 1170 \AA) radiation ionizes hydrogen. This process heats and expands the atmospheric layers above the planet's thermosphere, potentially leading to ion pickup by the stellar wind or hydrodynamic outflow of hydrogen \citep{lammer2007,tian2008, murrayclay2009, koskinen2010, rahmati2014, chadney2015,tripathi2015}. Depending on the amount of EUV flux they are exposed to, habitable zone planets around M stars can lose both oceans and significant fractions of their atmospheres within a few billion years \citep{luger2015}. Correct estimates of stellar EUV flux are important for studying the stability of exoplanet atmospheres and the stability of the M star habitable zone.

Observing in EUV wavelengths is extremely difficult due to optically thick interstellar hydrogen absorbing most of the spectrum between 400 -- 912 \AA \ \citep{barstow2007}. While the quantity of absorption due to the interstellar medium (ISM) is dependent on the direction, short-wavelength UV observations are significantly affected for nearly all planet-hosting stars ($>$100 pc) making them possible only for systems closer than 50 -- 100 pc \citep{fossati2017}. Over the course of the mission, the \textit{Extreme Ultraviolet Explorer} (\textit{EUVE}) observed six active M stars with low signal-to-noise from 100 -- 400 \AA \ (AD Leo, AU Mic, EV~Lac, Proxima Centauri, YY Gem, YZ CMI). The majority of the EUV emission from these stars is below the minimum detectable flux level for the instrument, so only strong emission lines including \ion{He}{2} and highly ionized iron (Fe {\small\rmfamily IX -- XVI \relax}) were identified \citep{craig1997}. No flux was observed from 350-912 \AA. While current capabilities allow for FUV, NUV, and limited X-ray measurements, there are no operational instruments able to observe stars other than the Sun in the EUV wavelength range. 

Due to the scarcity of EUV observations for M stars, studies have relied on using either the \textit{EUVE} spectrum of the very active M star, AD Leo \citep{segura2010, wordsworth2010}, or other proxies to determine the flux in this wavelength regime. Current methods to predict EUV flux include various empirical scaling relationships \citep{linsky2014,chadney2015,france2018} and semiempirical models to produce synthetic EUV spectra (e.g. \citealt{lecvaelier2007, sanzforcada2011, fontenla2016, peacock2019}). Stellar EUV spectra are characterized by many emission lines with large dynamic ranges in flux, superimposed on a continuum marked with distinct bound-free edges from continuous opacity sources. In a low activity solar spectrum, the strongest EUV emission lines peak at fluxes three orders of magnitude larger than continuum levels \citep{tobiska1996}. This level of detail is not encompassed in broadband scaling relationships and can only be predicted with high resolution synthetic spectra. The radiation in individual EUV emission lines penetrates planetary atmospheres at different depths, affecting the ionization rates and likelihood for escape. It is therefore important to include high resolution stellar EUV spectra rather than single valued fluxes when modeling both the photochemistry and escape in exoplanet atmospheres.

There has been significant effort to produce semiempirical models for the Sun (e.g. \citealt{vernazza1981,fontenla1990}) that utilize the wealth of solar observations across all wavelengths to validate the temperature structure. The flux in different wavelength regions emerges from different depths in a stellar atmosphere, providing important constraints on the thermal structure. Early modeling of M type stars used ground-based observations of chromospheric lines found in optical wavelengths to determine the temperature structure in the chromosphere (e.g. \citealt{giampapa1982,andretta1997,short1998,fuhrmeister2005})\footnote{For a complete summary of previous semiempirical chromosphere models for Main Sequence stars, see \cite{linsky2017}.}. In order to determine the temperature structure in the full upper atmosphere and estimate realistic EUV spectra, X-ray and/or UV observations are needed to guide and validate semiempirical stellar models. It is important to note that the timescales for flares and magnetic activity cycles occurring in the upper-atmospheric layers of M stars ranges from seconds to years and results in highly variable levels of measured X-ray and UV flux \citep{fossi1996,hawley2003,stelzer2013,loyd2018a,loyd2018b}. As a result, models based on non-contemporaneous X-ray and UV observations will have potentially large uncertainties in the predicted EUV spectra.

Stellar X-ray measurements provide important information about the thermal structure in the outermost $\sim$~10$^6$~K coronal layers, where emission features of highly ionized species found in the XUV (1 -- 912 \AA) spectrum form. The EUV continuum and many FUV emission lines, however, form at cooler temperatures deeper in the stellar atmosphere in the chromosphere and transition region. Estimating UV spectra using only X-ray observations leads to systematically underpredicted line fluxes because the X-ray observations do not include the contribution from the deeper atmospheric layers. For example, \cite{sanzforcada2011} used emission measure distribution coronal models with \textit{XMM-Newton}, \textit{Chandra}, and \textit{ROSAT} X-ray observations to compute synthetic XUV spectra of 82 late-F to mid-M planet hosts, but the spectra were found to underpredict observed FUV line strengths by up to a factor of 33 \citep{france2016,louden2017}. 

The majority of spectral features observed in the stellar UV spectrum form at temperatures ranging from 10$^4$ -- 10$^6$ K \citep{sim2005}. While FUV and NUV observations lack major contribution from the corona (with notable exception of the \ion{Fe}{12} line at 1242 \AA \ that forms near 1.4 $\times$ 10$^6$ K), they provide crucial information about the thermal structure in the chromosphere and transition region, where a large fraction of EUV emission is generated. Recently, \cite{peacock2019} used the PHOENIX atmosphere code to predict the EUV through near-IR spectrum of the M8 star TRAPPIST-1, calibrating the models with two UV datasets. The models have temperature-pressure profiles qualitatively similar to the Sun, but without a corona. The models reproduce the UV observations well and predict EUV fluxes consistent with estimates calculated using empirical scaling relationships. However, as a result of not including the coronal flux contribution, they are estimated to underpredict the spectrum at wavelengths $<$300 \AA. \cite{fontenla2016} adapted their solar model \citep{fontenla2007,fontenla2009,fontenla2011,fontenla2014,fontenla2015} to produce a full upper atmosphere model for the M1 star GJ 832 that covers X-ray through optical wavelengths and provides the first spectrally resolved EUV prediction for this star. 

In this paper, we present 1D nonlocal thermodynamic equilibrium (non-LTE) synthetic spectra (EUV -- IR, 100 \AA \ -- 5.5 $\mu$m) of three early-M planet host stars (GJ 832, GJ 176, GJ 436) that replicate both \textit{Hubble Space Telescope} (\textit{HST}) and \textit{Galaxy Evolution Explorer} (\textit{GALEX}) observations of each target. We show that constructing models with a simple temperature structure can reproduce the UV continuum and many emission features in high resolution \textit{HST} spectra. These models also predict the unobservable EUV spectrum, as they include prescriptions for the stellar upper atmosphere, including the chromosphere and transition region, where EUV, FUV, and NUV fluxes originate.

In Section \ref{sec:model}, we describe the construction and specifications in our atmospheric models. In Section \ref{sec:targets}, we discuss the properties and pre-existing observations of our target stars that we use to guide our models. We analyze the computed spectra and compare them to optical and UV observations in Section \ref{sec:compobs}. The synthetic EUV spectra are described in Section \ref{sec:euv}. In Section \ref{sec:discussion}, we compare our models to flux estimates derived from empirical scaling relationships and semiempirical modeling. Conclusions are given in Section \ref{sec:conclusions}.

\begin{figure}[t]
    \centering
    \includegraphics[width=1.0\linewidth]{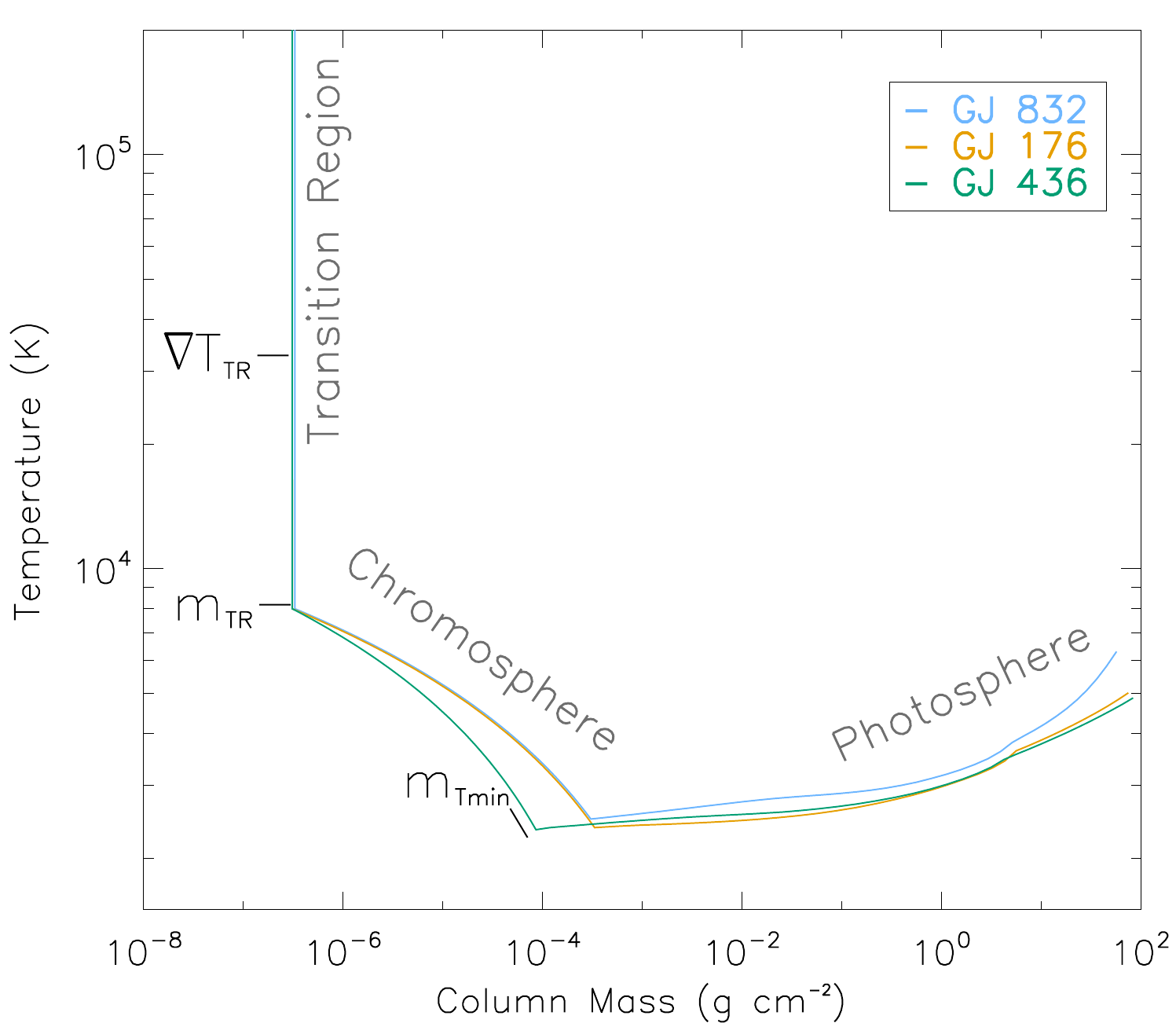}
    \caption{Temperature-column mass structures for models of GJ 832 (blue), GJ 176 (orange), and GJ 436 (green) with prescriptions for the chromosphere and transition region. The radiative-convective boundary in the photosphere for all three models occurs around 8 g cm$^{-2}$. Free parameters in the construction of the upper atmosphere are the column mass at the base and top of the chromosphere: m$_{Tmin}$ and m$_{TR}$, and the temperature gradient in the transition region: $\nabla$ $T_{TR}$ (approximate locations labeled). The parameter values for these models are given in Table \ref{tab:modelparam}.}
    \label{fig:Tstructure}
\end{figure}

\section{Model}\label{sec:model}

We build 1D stellar upper atmosphere models with prescriptions for the chromosphere and transition region using the multi-level non-LTE code PHOENIX \citep{hauschildt1993,hauschildt2006,baron2007}. PHOENIX has been used to study the diagnostic properties of strong chromospheric lines in the optical spectrum of M stars \citep{hauschildt1996,andretta1997, short1998, fuhrmeister2005, fuhrmeister2006, hintz2019} and to create EUV -- IR spectra of the M8 star, TRAPPIST-1 \citep{peacock2019}. 

For our models, we use a similar prescription to that used in \cite{peacock2019}. We begin with a base photosphere model in radiative-convective equilibrium that corresponds to the effective temperature, surface gravity, and mass of each star. We then superimpose an increasing temperature distribution up to 8,000~K to simulate a chromosphere and a steep temperature gradient above 8,000~K to simulate the transition region (Figure \ref{fig:Tstructure}).  At temperatures greater than $\sim$8,000~K, neutral hydrogen, the major source of electrons, is severely depleted by ionization and is no longer an efficient cooling agent. If heating decreases outward more slowly than column mass, then the chromosphere becomes thermally unstable, leading to the onset of the transition region. Stability is reattained near coronal temperatures where heat conduction and stellar wind losses dominate the plasma energy balance \citep{ayres1979}. In our models, we set the hottest layer at the top of the transition region to be 200,000 K, since the majority of observed emission lines in M star UV spectra form at or below this temperature, e.g. \ion{Mg}{2}~($\sim$10$^{4.2}$~K), \ion{C}{2}~(10$^{4.3}$~K), \ion{H}{1}~($\sim$10$^{4.5}$~K), \ion{Si}{4}~(10$^{4.78}$~K), \ion{He}{2}~($\sim$10$^{4.9}$~K), \ion{C}{4}~(10$^{4.98}$~K), \ion{N}{5}~(10$^{5.22}$~K) \citep{sim2005}. 

\begin{deluxetable}{l c c c c}[t]
\tablecaption{Model Parameters \label{tab:modelparam}}
\tablehead{
\colhead{Model}  &\colhead{$\nabla$T$_{TR}$} & \colhead{$m_{TR}$} & \colhead{$m_{Tmin}$} & \colhead{P$_{out}$}\\
\colhead{}  & \colhead{(K dyne$^{-1}$ cm$^2$)} & \colhead{(g cm$^{-2}$)} & \colhead{(g cm$^{-2}$)} & \colhead{(dyne cm$^{-2}$)}}
\startdata   
    GJ 832  & 10$^9$ & 10$^{-6.5}$ & 10$^{-3.5}$ & 0.016\\
    GJ 176  & 10$^9$ & 10$^{-6.5}$ & 10$^{-3.5}$ & 0.02\\
    GJ 436  & 10$^9$ & 10$^{-6.5}$ & 10$^{-4}$ & 0.02\\
\enddata
\end{deluxetable}

\begin{deluxetable*}{lccccccccc}[t!]
\tablecaption{Species computed in non-LTE\label{tab:nlte}}
\tablehead{
\colhead{Element} & \colhead{Abundance} & \multicolumn6c{Levels} & \colhead{Total Lines} \\
 &  & \colhead{I} & \colhead{II} & \colhead{III} & \colhead{IV} & \colhead{V} & \colhead{VI} & }
\startdata  
        H  & 12.0 & 30 & $\cdots$ & $\cdots$ & $\cdots$ &$\cdots$ &$\cdots$ & 435\\ 
        He  & 10.93 & 19 & 10 & $\cdots$ & $\cdots$ &$\cdots$ &$\cdots$ & 82\\ 
        C  & 8.43 & 230 & 85 & 79 & 36 &$\cdots$ &$\cdots$ & 4,216\\ 
        N  & 7.83 & 254 & 152 & 87 & 81 &$\cdots$ &$\cdots$ & 5,537\\ 
        O  & 8.69 & 146 & 171 & 137 & 138 &$\cdots$ &$\cdots$ & 3,466\\ 
        Ne  & 7.93 & 26 & 279 & $\cdots$ & $\cdots$ &$\cdots$ &$\cdots$ & 3,516\\ 
        Na  & 6.24 & 58 & 35 & 69 &$\cdots$ &$\cdots$ &$\cdots$ & 858\\
        Mg  & 7.6 & 179 & 74 & 90 & 54 &$\cdots$ &$\cdots$ & 2,972 \\ 
        Al  & 6.45 & 115 & 191 & 58 &$\cdots$ &$\cdots$ &$\cdots$ & 3,432\\
        Si  & 7.51 & 330 & 93 & 163 & 52 &$\cdots$ &$\cdots$ & 6,001\\
        P  & 5.41 & 230 & 90 & $\cdots$&$\cdots$ &$\cdots$ &$\cdots$ & 1,827\\
        S  & 7.12 & 152 & 84 & 41 &$\cdots$ &$\cdots$ &$\cdots$ & 2,666\\
        Cl & 5.5 & 278 & 124 & 78 &$\cdots$ &$\cdots$ &$\cdots$ & 8,314\\
        Ar & 6.4 & 208 & 302 & 96 &$\cdots$ &$\cdots$ &$\cdots$ & 7,652\\
        K & 5.08 & 80 & 22 & 38 &$\cdots$ &$\cdots$ &$\cdots$ & 834\\
        Ca & 6.34 & 196 & 89 & 150 &$\cdots$ &$\cdots$ &$\cdots$ & 4,730\\
        Ti & 4.95 & 555 & 204 & 185 & 39 &$\cdots$ &$\cdots$ & 18,317\\
        V & 3.93 & 483 & 323 & 299 &$\cdots$ &$\cdots$ &$\cdots$ & 17,246\\
        Cr & 5.64 & 392 & 733 & 214 &$\cdots$ &$\cdots$ &$\cdots$ & 23,464\\
        Mn & 5.43 & 297 & 512 & 391 &$\cdots$ &$\cdots$ &$\cdots$ & 17,214 \\
        Fe & 7.5 & 902 & 894 & 555 & 276 & 180 & 93 & 62,791\\
        Co & 4.99 & 364 & 255 & 213 &$\cdots$ &$\cdots$ &$\cdots$ & 11,548\\
        Ni & 6.22 & 180 & 670 & 344 &$\cdots$ &$\cdots$ &$\cdots$ & 26,063 
\enddata
\end{deluxetable*}

Heating mechanisms in M dwarf chromospheres are not well understood. Acoustic heating, magnetic heating, and back irradiation from coronal layers are all suggested potential contributing processes \citep{narain1996}. For our models, we assume a linear temperature rise with log(column mass) in both the chromosphere and transition region. Previous works have experimented with implementing nonlinear log(column mass)-temperature profiles and found that nonlinear temperature rises can be tailored to replicate individual lines very well, but linear rises give the best overall continuum fit \citep{eriksson1983,andretta1997,fuhrmeister2005}. For example, \cite{fuhrmeister2005} altered the structure of the lower chromosphere in order to better fit observations of \ion{Na}{1} D lines, but found that it resulted in worse fit models with significantly increased flux in the Balmer lines.

In the construction of the upper atmosphere, we alter three free parameters designating the depth at which the upper atmosphere is attached to the underlying photosphere and the thickness of both the chromosphere and transition region (approximate locations labeled in Figure \ref{fig:Tstructure}). The specific parameters are the column mass at the initial chromospheric temperature rise ($m_{Tmin}$), the column mass at the top of the chromosphere ($m_{TR}$), and the temperature gradient in the transition region ($\nabla$~$T_{TR}$~=~$|$ d~$T$/d~log~$P|$). We then use the prescriptions selected for each model to calculate the outer pressure:
\begin{equation}
    P_{out} = \frac{T_{max} - T_{ch}}{\nabla T_{TR}} + (m_{TR}*g)
\end{equation}

where T$_{max}$ is the temperature at the top of the transition region, set to 200,000 K, T$_{ch}$ is the temperature at the base of the transition region, set to 8,000~K, and $g$ is the surface gravity of the star. For each star, we create a grid of 29 models varying $m_{Tmin}$ = 10$^{-6}$ -- 10$^{-3.5}$ g cm$^{-2}$, $m_{TR}$ = 10$^{-7}$ -- 10$^{-6}$ g cm$^{-2}$, and $\nabla$~$T_{TR}$~= 10$^{8}$ -- 10$^{10}$ K dyne $^{-1}$ cm$^2$. 

Different states of stellar activity are simulated by adjusting where the chromospheric temperature rise is attached to the underlying photosphere model \citep{andrgiam1995, andretta1997}. Higher activity states, and therefore, higher UV continuum fluxes, are generated by shifting the temperature structure uniformly inward, beginning the initial temperature rise at larger column mass. Since many FUV emission features form at similar temperatures to those that are found in the EUV spectrum \citep{kretz2009}, both wavelength regimes are most sensitive to the same parameters: $\nabla T_{TR}$ and $m_{TR}$. A factor of 10 decrease in $\nabla T_{TR}$ or increase in $m_{TR}$ results in an order of magnitude increase in the integrated flux density across EUV wavelengths and a factor of five increase in the integrated FUV flux density. Changes in $m_{Tmin}$ correspond to changes in the strength of the extended wings of Ly$\alpha$ and therefore the FUV pseudo-continuum, but have a larger effect on the NUV spectrum. Adjusting the depth of the chromosphere by shifting the position where the temperature rise begins inwards by 10 g cm$^{-2}$ increases the NUV flux density by a factor of three and the FUV flux density by up to a factor of two. Changes in $m_{Tmin}$ by itself have no effect on the computed EUV spectrum.

\subsection{Microturbulent Velocity}

Levels of microturbulent velocity ($v_{turb}$) influence the intensity and wing shape of emission lines, particularly affecting hydrogen, sodium, and calcium lines \citep{jevremovic2000}. The turbulent surface of a stellar photosphere induces disturbances in the overlying chromosphere that are propagated at the speed of sound. For our models, we set $v_{turb}$ to 2 km s$^{-1}$ in the photosphere and 10 km s $^{-1}$ at the top of the transition region. We follow the same approach as \cite{fuhrmeister2005} in setting the slope of $v_{turb}$ in the chromosphere and lower transition region to a fraction of the sound speed in each layer, but not allowing this value to exceed 10~km~s~$^{-1}$. We select a value of 0.35~$\times$~$v_{sound}$ such that $v_{turb}$ smoothly transitions from 2~km~s$^{-1}$ in the photosphere to the larger values in the chromosphere.

\subsection{Non-Local Thermodynamic Equilibrium  Radiative Transfer}	

In an M dwarf photosphere, temperatures are low and densities are high, such that collisions dominate and LTE is an appropriate approximation for calculating the level populations of atoms and molecules. As temperatures increase in the upper atmosphere and densities become very low, radiative rates exceed collisional rates and radiative transfer is dominated by non-LTE effects. PHOENIX is equipped with current atomic level data \citep{dere1997,kurucz2014,kurucz2017,delzanna2015} suitable for these high temperatures and low densities and has the capacity to do multi-line non-LTE calculations for many species. For our models, we consider a total of 15,355 levels and 233,871 emission lines when computing the set of 73 atoms and ions listed in Table ~\ref{tab:nlte} in full non-LTE radiative transfer using species and background opacities provided by the PHOENIX and CHIANTI v8 databases \citep{Landi2006}. Also included in our models are new bound-free molecular opacities described in \cite{peacock2019} and millions of optically thin atomic and molecular background lines calculated with assuming an LTE source function.
	
\subsection{Partial Frequency Redistribution}

Strong resonance lines are typically good indicators of chromospheric activity. They are characterized as optically thick lines with broad absorption wings that form in the photosphere and lower chromosphere with cores that form in the upper chromosphere and transition region. Since these lines dictate information regarding the thermodynamic properties of the stellar upper atmosphere and provide constraints on the Lyman continuum and EUV spectrum \citep{linsky2014,shkolnik2014}, it is important to model these lines particularly well. 

\begin{figure}
    \centering
    \includegraphics[width=1.0\linewidth]{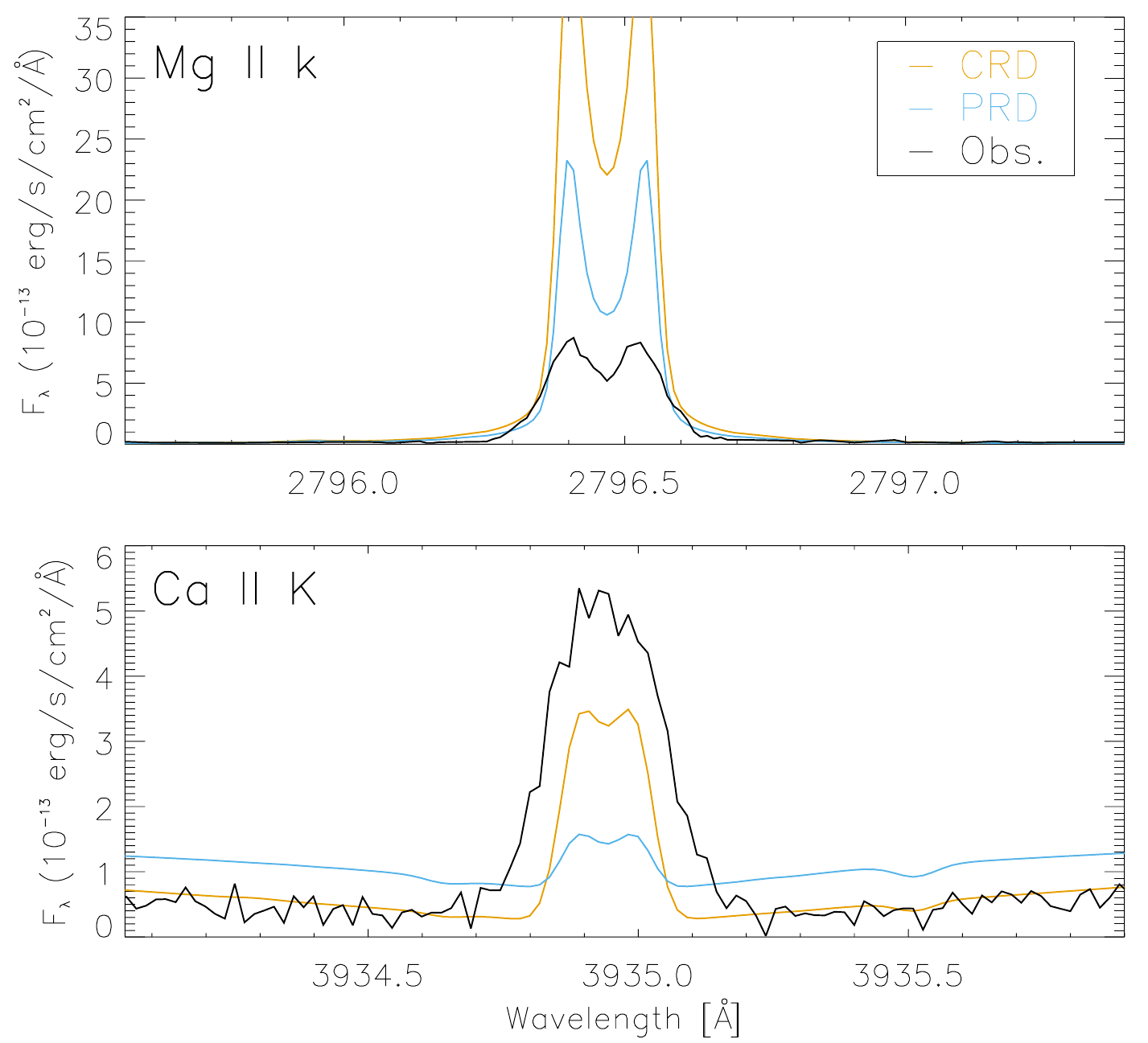}
    \caption{Computed non-LTE profiles for GJ 832 assuming CRD (orange) versus PRD (blue) for \ion{Mg}{2} \textit{k} (top panel) and \ion{Ca}{2} K (bottom panel) compared to high resolution observed spectra (black; top panel: \textit{HST} STIS \citep{france2016}, bottom: Keck/HIRES \citep{vogt2011}).}
    \label{fig:PRD}
\end{figure}

Complete frequency redistribution (CRD) accounts for overlapping radiative transitions and is generally a good approximation to use when calculating the majority of line profiles. In strong resonance lines, however, coherent scattering of photons largely affects the shape of the wings in the line profiles, requiring the inclusion of partial frequency redistribution (PRD) in the radiative transfer calculations. 

As a part of \cite{peacock2019}, we added PRD capabilities to PHOENIX and demonstrated the importance of these calculations when computing the \ion{H}{1} Ly$\alpha$ line in M stars. In this paper, we extend these calculations to additional strong resonance lines that are commonly used as chromospheric diagnostics: \ion{Mg}{2} \textit{h} \& \textit{k} and \ion{Ca}{2} H \& K. In Figure \ref{fig:PRD}, we show the impact of including the PRD formalism when computing the \ion{Mg}{2} \textit{k} and \ion{Ca}{2} K line profiles in our model for GJ~832. When computed assuming CRD, the model \ion{Mg}{2} \textit{k} profile marginally overpredicts the wings and overpredicts the strength of the observed line core by a factor of $\sim$13. Observations of \ion{Mg}{2}, however, are contaminated by interstellar absorption, so direct comparisons to the line core cannot be drawn. Estimates suggest $\sim$30\% attenuation of the intrinsic \ion{Mg}{2} line core, affecting the line profiles from 2796.4 -- 2796.6 \AA \ and 2803.6 -- 2803.7 \AA \citep{france2013} as compared to the nearly 100\% interstellar absorption of the \ion{H}{1} Ly$\alpha$ core \citep{youngblood2016}. We find that the PRD calculations decrease the flux in the \ion{Mg}{2} line peaks and line core by nearly a factor of two, with marginal effects on the strength of the wings.

The CRD \ion{Ca}{2} K profile slightly underpredicts the observed profile, with the total line flux differing by a factor of 1.9. Computing \ion{Ca}{2} in PRD increases the strength of the wings and decreases the total line flux by a factor of four. Since the PRD calculations worsen the agreement to the observations for \ion{Ca}{2}, we use the CRD formalism in computing this species in our models.

The broad wings of Ly$\alpha$ extend far enough from the line center that using PRD to compute this line profile drastically alters the FUV pseudocontinuum and influences which set of model parameters produces the spectrum that most closely matches the observations. Conversely, computing \ion{Mg}{2} and \ion{Ca}{2} in PRD versus CRD results in negligible changes to the surrounding continuum, such that the special treatment of these lines does not have an effect on the choice of model that best reproduces the observations as a whole, and therefore does not effect the predicted EUV spectrum. In future work, we will explore a non-linear temperature rise in the chromosphere, in which the accuracy in modeling both \ion{Mg}{2} and \ion{Ca}{2} will play a more vital role in determining the structure.

\begin{figure*}[t!]
    \centering
    \includegraphics[scale=1.1]{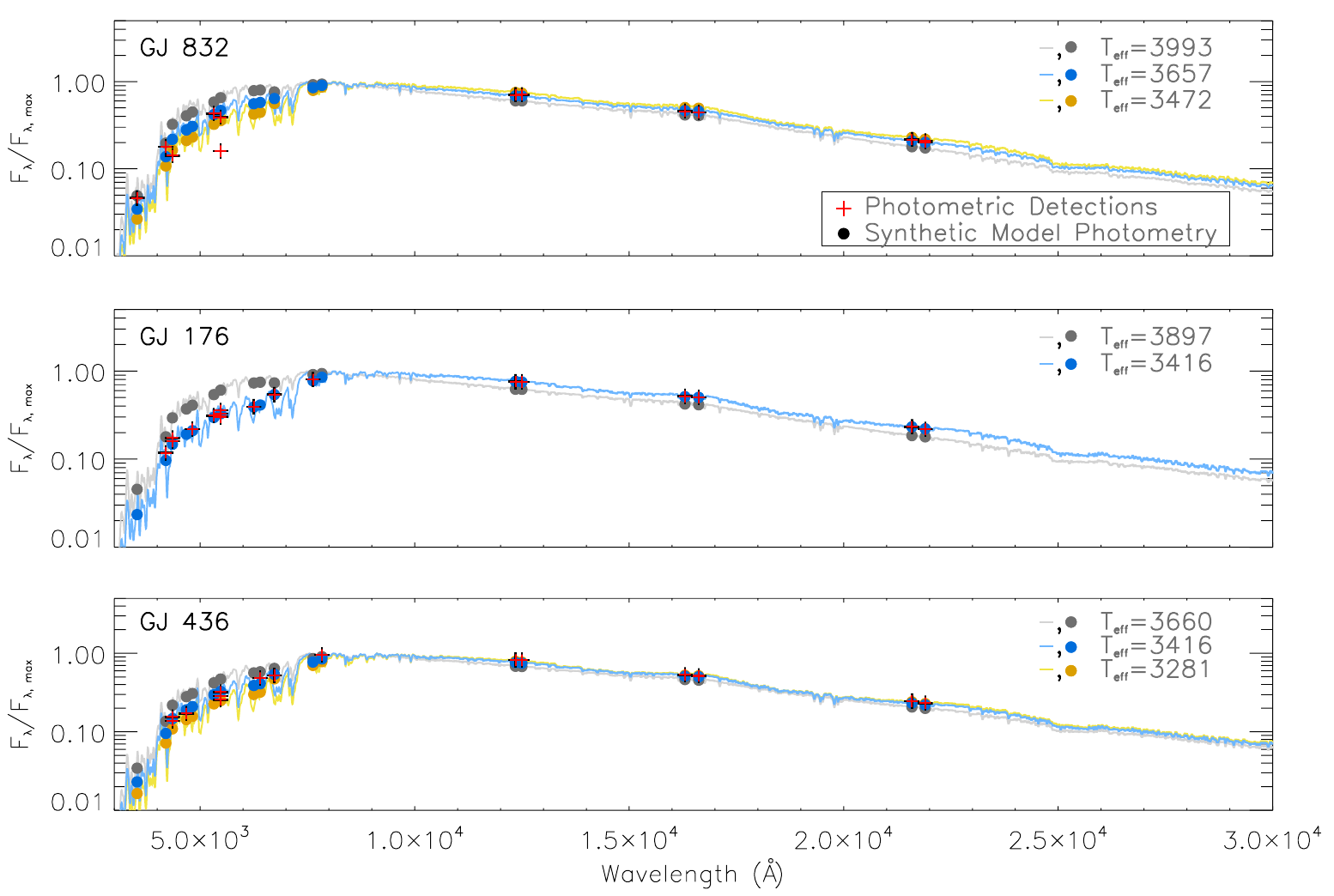}
    \caption{Flux-normalized PHOENIX synthetic spectra and synthetic photometry (circles) for the range of effective temperatures found in the literature (coldest: yellow, hottest: grey) compared to all available visible and near-IR photometric detections per star returned by the VizieR Photometry Viewer (red crosses). Best fit models are plotted in blue. Spectral resolution in the spectra has been degraded for clarity. \\
    References for photometric detections -- \\
    GJ832: 1,2,3,4,5,6,7,8,9,10,11;  GJ176: 1,2,3,4,5,6,8,9,10,11,12,13,14,15,16,17,18; GJ436: 1,2,3,5,8,9,10,12,13,15,16,19,20\\
    References -- (1) \citealt{cutri2003}; (2) \citealt{roser2008}; (3) \citealt{roeser2010}; (4) \citealt{zacharias2012}; (5) \citealt{neves2013}; (6) \citealt{santos2013}; (7) \citealt{bourges2014}; (8) \citealt{gaidos2014}; (9) \citealt{wardduong2015}; (10) \citealt{altmann2017}; (11) \citealt{schneider2018}; (12) \citealt{soubiran2016};  (13) \citealt{droege2006}; (14) \citealt{huber2017}; (15) \citealt{trifonov2018}; (16) \citealt{terrien2015}; (17) \citealt{ofek2008}; (18) \citealt{ammons2006}; (19) \citealt{lasker2008}; (20) \citealt{triaud2014}}
    \label{fig:visIRspec}
\end{figure*}	
	
\begin{deluxetable*}{lccc}[t!]
\tablecaption{Stellar Parameters \label{tab:stelparam}} 
\tablehead{
\colhead{} & \colhead{GJ 832} & \colhead{GJ 176} & \colhead{GJ 436}
}
\startdata
        Spectral Type\dotfill             & M1.5\tablenotemark{1} & M2.5\tablenotemark{2} & M3.5\tablenotemark{3}\\
        Age (Gyr)\dotfill                 & $>$3.1\tablenotemark{4} & 3.62\tablenotemark{5} & 6$^{+4}_{-5}$ \tablenotemark{6} \\
        T$_{eff}$ (K)\dotfill             & 3657\tablenotemark{1} & 3416 $\pm$ 100\tablenotemark{7} & 3416 $\pm$ 100\tablenotemark{8} \\
        log(g) (cm s$^{-2}$)\dotfill  & 4.7\tablenotemark{9} & 4.79 $\pm$ 0.13\tablenotemark{10} & 4.84 $\pm$ 0.018\tablenotemark{6} \\
        M$_{\star}$ (M$_{\odot}$)\dotfill & 0.45 $\pm$ 0.05\tablenotemark{9} & 0.45\tablenotemark{11} & 0.507$^{+0.071}_{-0.062}$\tablenotemark{  3}\\
        R$_{\star}$ (R$_{\odot}$)\dotfill & 0.499 $\pm$ 0.017\tablenotemark{12} & 0.45 $\pm$ 0.02 \tablenotemark{11} & 0.455 $\pm$ 0.018\tablenotemark{11} \\
        $[Fe/H]$\dotfill                  & -0.17 $\pm$ 0.09\tablenotemark{13} & -0.01 $\pm$ 0.09\tablenotemark{13} & -0.03 $\pm$ 0.09\tablenotemark{13} \\
        $v$sin(i) (km s$^{-1}$) \dotfill                   & $\cdots$ & $<$0.8\tablenotemark{2} & 0.24$^{+0.38}_{-0.17}$\tablenotemark{   8}\\
        $v_{rad}$ (km s$^{-1}$)\dotfill   & 12.52 $\pm$ 0.15\tablenotemark{14} & 25.68 $\pm$ 0.29\tablenotemark{14} & 9.6 \\
        Distance (pc)\dotfill             & 4.965 $\pm$ 0.001\tablenotemark{14} & 9.473 $\pm$ 0.006\tablenotemark{14} & 9.755 $\pm$ 0.007\tablenotemark{14} \\
\enddata
\tablerefs{(1) \citealt{bailey2009}; (2) \citealt{forveille2009}; (3) \citealt{vonbraun2014}; (4) \citealt{newton2016}; (5) \citealt{sanzforcada2011}; (6) \citealt{torres2007}; (7) \citealt{loyd2016}; (8) \citealt{lanotte2014}; (9) \citealt{wittenmyer2014}; (10) \citealt{santos2013}; (11) \citealt{vonbraun2014}; (12) \citealt{houdebine2010}; (13) \citealt{neves2014}; (14) \citealt{gaia2018}}
\end{deluxetable*}		
	
\section{Targets}\label{sec:targets}

There are few M stars that have observations in X-ray, UV, optical, and IR wavelengths. For this work, we modeled 3 M star planet hosts of different subtypes that have UV spectroscopic observations and estimates for EUV fluxes calculated from empirical scaling relationships with Ly$\alpha$ from the MUSCLES Treasury Survey program \citep{france2016,youngblood2016,loyd2016} as well as \textit{GALEX} UV photometric detections \citep{bianchi2011,shkolnik2014}: 

\textit{GJ 832} is an M1.5 V star that hosts two planets: a 0.64 M$_{Jupiter}$ planet at a semi-major axis of 3.56 au and a 5.4 M$_{\oplus}$ super-Earth located in the canonical habitable zone (0.16 au) \citep{wittenmyer2014}. Its age is uncertain, but a rotational period of 45.7 $\pm$ 9.3 days \citep{suarez2015} indicates that it has a kinematic age of at least 3.1 Gyr \citep{newton2016, west2015}. GJ~832 has also been previously modeled in \cite{fontenla2016}, allowing for the opportunity to directly compare the two synthetic EUV spectra.

\textit{GJ 176} is a 3.62 Gyr \citep{sanzforcada2011} M2.5~V star that hosts an 8.3 M$_{\oplus}$ planet orbiting at 0.066 au \citep{butler2009, forveille2009}. \cite{youngblood2016} calculated the column density of \ion{H}{1} in the interstellar medium along the line of sight to GJ 176 and found the value to be among the lowest column densities measured in their sample of 11 M and K dwarfs. In the event of a new EUV telescope, the smaller concentration of interstellar hydrogen in the direction of GJ 176 suggests that it would be a favorable target for follow-up EUV observations.

\textit{GJ 436} is a 6$_{ -5}^{+4}$ Gyr \citep{torres2007} M3.5 V. Ly$\alpha$ transit observations of its Neptune-sized planet (23 M$_{\oplus}$, 0.287 au) discovered by \citealt{butler2004} show a 56\% transit depth that is most likely caused by a large cloud of escaping hydrogen \citep{kulow2014,ehrenreich2015}. Since EUV radiation drives atmospheric escape, a synthetic stellar EUV spectrum is essential for understanding the star-planet interaction in this system.

The effective temperature (T$_{eff}$) is a primary driver of spectral shape, especially affecting NUV and visible wavelengths in M stars. There is significant variance in the literature values of T$_{eff}$ for each target in our sample. Effective temperatures for GJ 832 range from 3472 K \citep{wittenmyer2014} -- 3993 K \citep{gaia2018}, for GJ 176 range from 3416 K \citep{loyd2016} -- 3897 K \citep{gaia2018}, and for GJ 436 range from 3281 K \citep{loyd2016} -- 3660 K \citep{gaia2018}. We computed models with the range of literature values and calculated synthetic visible and near-IR photometry over the same wavelengths as the filter profiles for: HIP: BT, VT; Johnson: B, V, J, H, K; POSS-II: J, F, i; Gaia:~G; 2MASS: J, H, Ks, SDSS: g, r, i. We normalized the fluxes and compared the values to all available photometric detections (not including upper limits) returned by the VizieR Photometry Viewer\footnote{http://vizier.u-strasbg.fr/vizier/sed/} for each star within a 5" search radius. We used a $\chi^2$ test to identify the best fit models to the photometry and to select our operational T$_{eff}$ for each star. The best fit models are shown in blue in Figure \ref{fig:visIRspec}, along with the upper and lower end members. Our operational stellar parameters are listed in Table ~\ref{tab:stelparam}.

\section{Comparison of Models to Observations}\label{sec:compobs}

We calibrate our models to replicate \textit{HST} COS and STIS observations taken as a part of the MUSCLES Treasury Survey (Version 2.1) \cite{france2016} and Keck High Resolution Echelle Spectrometer (HIRES) observations from \cite{vogt2011}. We scale the high resolution ($\Delta\lambda\sim$0.01 \AA) model spectra by $R_{\star}^2/d^2$ and account for radial velocity shifts, rotational broadening, and instrumental broadening before comparing to observations.

We determine the steepness of the temperature gradient in the transition region, $\nabla T_{TR}$, by fitting observations of emission lines with formation temperatures greater than 8,000~K (\ion{Si}{4}, \ion{He}{2}, \ion{C}{4}, and \ion{N}{5}). We use the FUV and NUV psuedocontinuum and the wings of Ly$\alpha$, \ion{Mg}{2} \textit{h} \& \textit{k}, and \ion{Ca}{2} H \& K, which have formation temperatures below 8,000~K, to determine $m_{Tmin}$ and $m_{TR}$. Best fits were determined by eye and with a $\chi^2$ test using the mentioned diagnostic lines. As we are using a 1D simplified linear temperature structure to model a 3D object with spatially varying active regions, it is not expected that the models will reproduce all observed chromospheric lines well. A good match to the observed UV continuum and many emission lines that cover a broad range of formation temperatures is indicative that the simplified upper atmospheric temperature structure and predicted spectrum are a good general approximation for the star as a whole. The model parameters that best replicate the \textit{HST} observations are given in Table \ref{tab:modelparam}. 

\begin{figure*}[t!]
    \centering
    \includegraphics[scale=1.1]{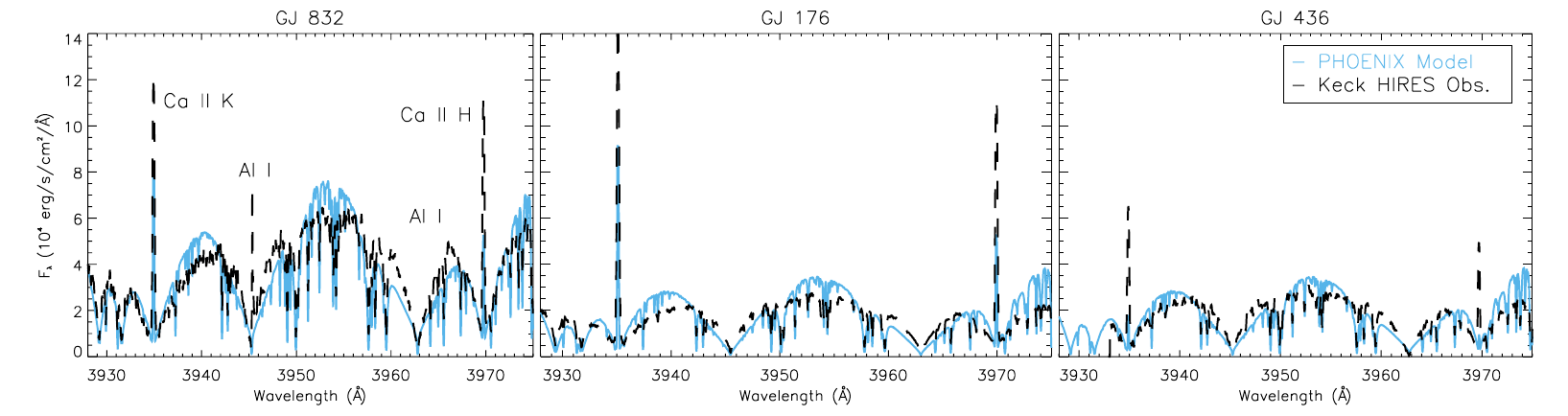}
    \caption{Comparison of the PHOENIX spectra (blue, solid) to \ion{Ca}{2} H (3968.17 \AA) \& K (3933.66 \AA) Keck/HIRES observations (black, dashed) \citep{vogt2011}. Observations are converted to vacuum wavelengths and scaled to the stellar surface. The resolution of the synthetic spectra has been reduced to that of the observations, $\Delta\lambda$ = 0.05 \AA.}
    \label{fig:ca2}
\end{figure*}

\subsection{Visible Spectrum}\label{subsec:vis}

The emission strengths of \ion{Ca}{2} H (3968.17 \AA) \& K (3933.66 \AA) and H$\alpha$ (6562.8 \AA) are commonly used indicators of chromospheric activity found in the visible region of a stellar spectrum. H$\alpha$ is a commonly used diagnostic for solar-type stars since the nearby continuum emission is relatively weak and it typically has broad wings. However, for M type stars, H$\alpha$ can be a complicated diagnostic of chromospheric activity since it progresses from presenting as an absorption feature to an emission feature with increasing activity \citep{gomes2011}. The \ion{Ca}{2} H \& K doublet does not undergo the transition from absorption to emission, making it a more favorable diagnostic feature detectable from the ground. For example, while the H$\alpha$ absorption spectra for the stars in this study imply that they are optically inactive \citep{gizis2002}, the strength of the \ion{Ca}{2} H \& K emission cores and observations of UV flares indicate that they do display chromospheric activity \citep{vogt2011,shkolnik2014,france2016,loyd2018b}. 

In M dwarfs, resonance lines of ions, including the \ion{Ca}{2} doublet, have very weak wings because the photosphere and lower chromosphere are mostly neutral. The line cores of \ion{Ca}{2} form in the upper chromosphere/lower transition region between 5,000 and 20,000 K while the neighboring emission peaks form at cooler temperatures deeper in the atmosphere between 4,000 and 10,000~K. Since these lines yield information about the temperature structure, before stellar UV spectral observations were available, early semiempirical M dwarf chromosphere models were based on fitting ground-based \ion{Ca}{2} K observations \citep{giampapa1982}. More recently, \cite{youngblood2017} found that the equivalent width of the \ion{Ca}{2} K line could be used to estimate the stellar surface flux in several ultraviolet emission lines, including Ly$\alpha$, and developed a scaling relationship to estimate EUV fluxes using time-averaged high resolution \ion{Ca}{2} observations.

We compare our model spectra to Keck/HIRES observations of \ion{Ca}{2} H \& K from \cite{vogt2011} in Figure \ref{fig:ca2}. Deviations of the synthetic spectra from the observations in the continuum surrounding the \ion{Ca}{2} doublet are related to the model parameter at the base of the chromosphere, $m_{Tmin}$. Adjusting $m_{Tmin}$ to better fit this region of the spectrum worsens the agreement of the model to the NUV continuum and does not influence the EUV spectrum. For all three models, the far wings of the lines replicate the observations well, but the models underpredict the observed emission cores by 45 -- 70\%. We attribute uncertainties in our estimated \ion{Ca}{2} lines to the lack of ambipolar diffusion in the model. Ambipolar diffusion is associated with coronal back-heating and is important for determining the hydrogen ionization near where the cores of \ion{Ca}{2} H \& K are forming ($\sim$ 10,000 K). In upcoming work, we will add a corona and ambipolar diffusion to our model and will explore how this physics affects the temperature at which hydrogen becomes fully ionized, which determines the onset of the transition region.

\subsection{Near Ultraviolet Spectrum}

\begin{figure} [t!]
    \centering
    \includegraphics[width=1.0\linewidth]{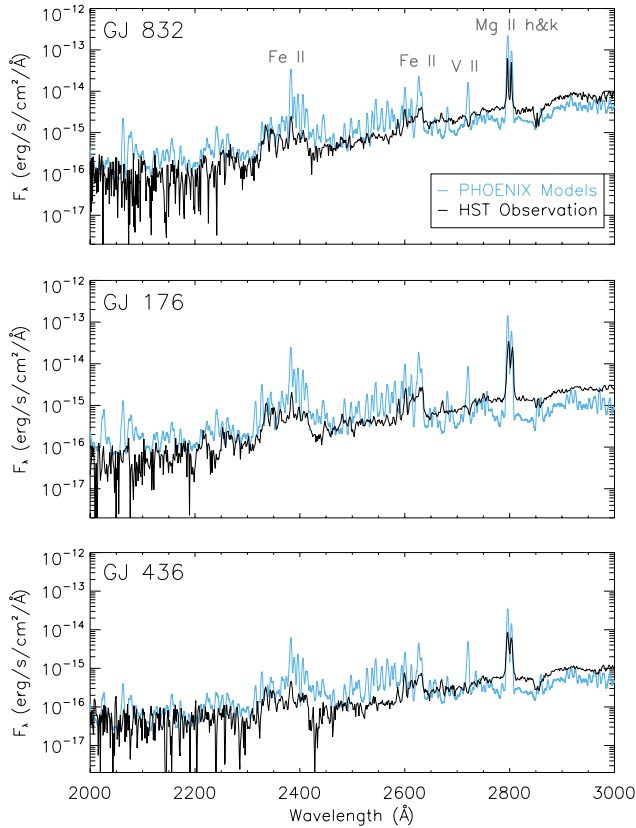}
    \caption{Near-UV PHOENIX spectra (blue) at the spectral resolution of the \textit{HST} observations (black). Prominent emission features are labeled in the top panel. The \textit{HST} observations for GJ 176 and GJ 436 were taken with the G230L grating on STIS, while the spectrum for GJ 832 is comprised of observations taken with the multiple gratings on both STIS and COS, including the high resolution grating used to measure \ion{Mg}{2}.}
    \label{fig:specnuv}
\end{figure}

At low resolution ($\Delta\lambda\sim$3 \AA), the NUV spectrum is characterized by the \ion{Mg}{2} \textit{h} (2802.7 \AA) \& \textit{k} (2795.53 \AA) doublet and a forest of \ion{Fe}{2} emission lines spanning 2300 -- 2650 \AA. The pseudo-continuum is shaped by molecular opacity sources, most importantly: NH, CH, OH, and H$_2$. In Figure \ref{fig:specnuv}, we compare our synthetic spectra to \textit{HST} observations taken as part of the MUSCLES Treasury survey. We find good general agreement in the pseudo-continuum for all three stars. From 2375~--~2630~\AA, the pseudo-continuum is elevated by excess flux from OH and H$_2$ causing the models to overpredict the forest of \ion{Fe}{2} lines. The models also overpredict a strong \ion{V}{2} line at 2721 \AA.

The \ion{Mg}{2} \textit{h} \& \textit{k} lines are diagnostics of the chromospheric thermal structure located in the NUV region. The line cores form at temperatures around 10,000 K, the emission peaks at $\sim$6,000 K, and the wings near 3,000 -- 4,000 K. This emission feature is similar to the \ion{Ca}{2} H \& K doublet in that both display weak wings, but the \ion{Mg}{2} lines typically have stronger emission cores in M dwarfs. This is due to a higher abundance of Mg than Ca, a larger ionization potential of Mg+ than Ca+, and a weaker background photospheric spectrum at shorter wavelengths \citep{linsky2017}.

Observations of \ion{Mg}{2} are contaminated by interstellar absorption, estimated to attenuate 30 -- 35\% of the intrinsic line flux \citep{france2013}. Additional absorption of \ion{Mg}{2} lines has been observed in host stars during the transit of exoplanets with escaping atmospheres, such as WASP-12b, attributed to metals in the exospheric cloud \citep{fossati2010}. It is therefore possible that the intrinsic \ion{Mg}{2} emission profiles of GJ 436 are further attenuated by the escaping atmosphere of its warm Neptune mass planet. Due to the low resolution modes used on the G230L grating on COS and STIS, corrections were not done for \ion{Mg}{2} for any of the \textit{HST} observations \citep{france2016}. Calculated line fluxes for \ion{Mg}{2} \textit{h} \& \textit{k} for the models and observations are given in Table \ref{tab:lineflux}. The spectral resolution of the models was reduced to that of the observations before calculating the line fluxes. We have also increased the observed line fluxes by 30\% in order to make a more accurate comparison to the intrinsic stellar values, and find that our model predictions are within a factor of 1.5 of the corrected \ion{Mg}{2} fluxes for GJ 176 and GJ 436, and within a factor of $\sim$7 for GJ 832.

\subsection{Far Ultraviolet Spectrum}

\begin{figure*}
    \centering
    \includegraphics[]{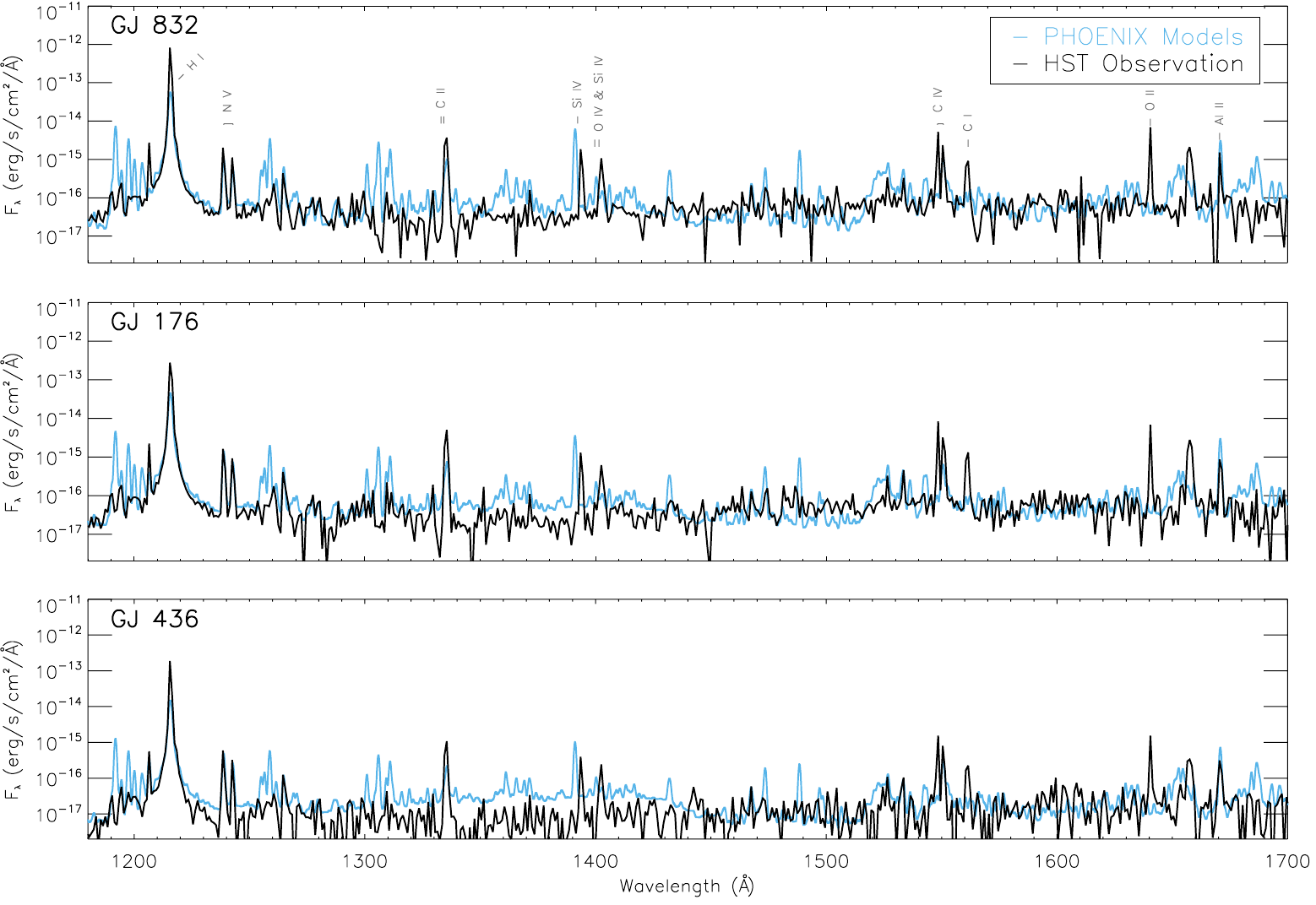}
    \caption{Comparison of the far-UV PHOENIX spectra (blue) to the MUSCLES composite \textit{HST} spectra (black). Both spectra are convolved to a resolution of 1 \AA. Prominent emission features are indicated in the top panel.}
    \label{fig:specfuv}
\end{figure*}

In Figure \ref{fig:specfuv}, we compare our synthetic FUV spectra to the MUSCLES composite \textit{HST} observations. We find good general agreement with the FUV continuum, which is shaped by recombination edges of Si, Mg, and Fe. 

\begin{deluxetable*}{cc|cc|cc|cc}[t!]
\tablecaption{Line fluxes (erg cm$^{-2}$ s$^{-1}$) of select chromosphere and transition region lines}
\tablehead{
 \colhead{Species} & \colhead{$\lambda$} & \multicolumn{2}{c}{GJ 832} & \multicolumn{2}{c}{GJ 176} & \multicolumn{2}{c}{GJ 436} \\
      & \colhead{(\AA)}        \vline &  
 \colhead{Model} & \colhead{\textit{HST}} \vline &
 \colhead{Model} & \colhead{\textit{HST}} \vline &
 \colhead{Model} & \colhead{\textit{HST}} 
 }
\startdata
    \ion{Si}{3} & 1206.51 &9.76&  24.76   & 17.05 & 40.24 & 45.25 & 49.87\\
    \ion{H}{1} (Ly$\alpha$)\tablenotemark{a} & 1215.67 &4033.81& 6602.65 & 1438.29 & 3656.67 & 2976.69 & 5906.28\\
    \ion{N}{5}& 1238.82 &63.08& 56.23   & 68.53 & 69.77& 110.13 & 46.47\\
    \ion{N}{5}& 1242.81 &34.61& 29.12   & 36.01 & 36.18 & 57.04 & 23.62\\
    \ion{C}{2}\tablenotemark{b}  &1334.53 &29.59& 26.63 & 5.55 & 22.97 & 5.06 & 9.04\\
    \ion{C}{2}\tablenotemark{b} & 1337.71 &59.24&  46.27 & 14.38 & 70.36  & 16.39 & 24.43\\
    \ion{Si}{4}& 1393.76 &5.64& 56.37  & 2.01 & 48.19 & 1.26 & 11.82\\
    \ion{O}{4}& 1401.16 &3.48& 5.11  & 1.34 & 3.92 & 1.41 & 1.51\\
    \ion{Si}{4}& 1402.77 &10.27& 31.4   & 2.45 & 23.04 & 2.42 & 7.27\\
    \ion{C}{4}& 1548.19 &9.87& 65.70   & 7.50 & 171.18     & 8.96 & 144.40\\
    \ion{C}{4}& 1550.77 &41.54& 32.87   & 25.38 & 89.63     & 42.93 & 69.86\\
    \ion{O}{2}& 1639.77 &5.45& 140.54  & 4.11 & 142.39     & 4.91 & 67.65\\
    \ion{Al}{2}& 1670.79 &247.25& 42.98   & 144.79 & 38.09    & 119.68 & 21.91\\
	\ion{Mg}{2} \textit{k}\tablenotemark{b}& 2796.35 &317.10 & 43.462 &203.84& 124.97 &126.26& 84.43\\
	\ion{Mg}{2} \textit{h}\tablenotemark{b}& 2803.53 &138.58 &  29.143 &84.05& 88.66 &88.5& 57.77
    \enddata
\tablecomments{Computed and observed emission line fluxes at the stellar surface and continuum normalized. \textit{HST} fluxes are calculated from the MUSCLES v2.1 panchromatic spectral energy distributions that maintain the native instrument resolutions and have been scaled by $d^2/R_{star}^2$.}
\tablenotetext{a}{Observed Ly$\alpha$ fluxes are calculated using the reconstructed profile from \cite{youngblood2016}.}
\tablenotetext{b}{Observed emission line fluxes have been corrected for an estimated 30\% absorption by interstellar \ion{Mg}{2} and \ion{C}{2}.}
\label{tab:lineflux}
\end{deluxetable*}

\begin{figure*}
    \centering
    \includegraphics[]{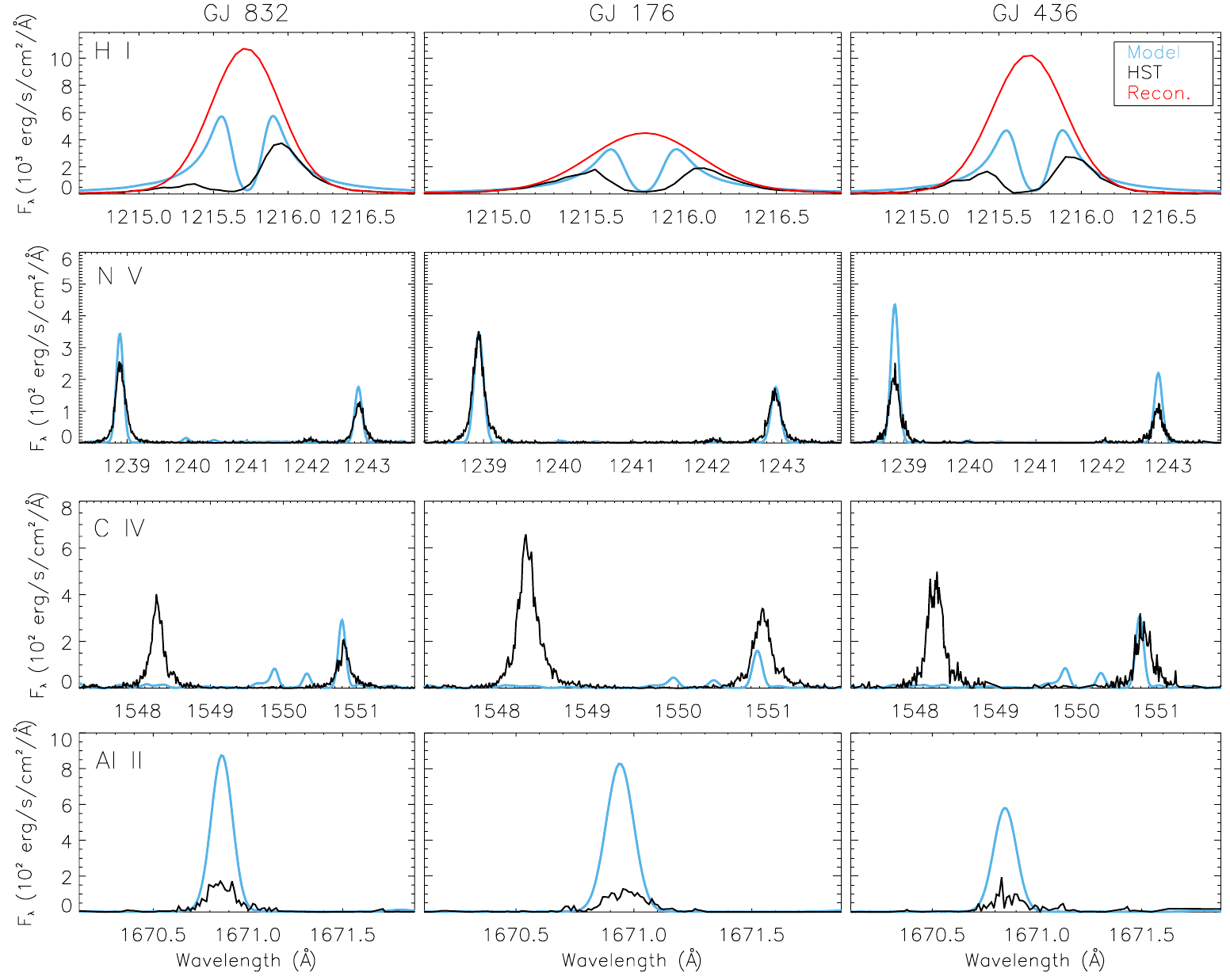}
    \caption{Select FUV emission line profiles from the PHOENIX spectra (blue) compared to \textit{HST} spectra (black) continuum normalized and scaled to the surface of the star. In the top panels, the raw \ion{H}{1} Ly$\alpha$ observation (black) is contaminated by interstellar absorption. The reconstructed profiles from \citealt{youngblood2016} are plotted in red. The central reversal in the model Ly$\alpha$ profiles are a result of non-LTE effects. The suppression of the \ion{C}{4} ($\sim$1548 \AA) line is a result of the linear parameterization in the transition region ($\nabla T_{TR}$, $m_{TR}$), as the \ion{C}{4} lines form at slightly different temperatures found near the top of the chromosphere and base of the transition region.}
    \label{fig:lines}
\end{figure*}

The FUV spectrum is populated with emission lines that form in both the chromosphere and transition region, including \ion{H}{1}, \ion{N}{5}, \ion{C}{1}, {\small\rmfamily II \relax} \& {\small\rmfamily IV \relax}, \ion{O}{2} \& {\small\rmfamily IV\relax}, and \ion{Si}{4} (labeled in the top panel of Figure \ref{fig:specfuv}). Individual emission line profiles for \ion{H}{1} Ly$\alpha$, \ion{N}{5}, \ion{C}{4}, and \ion{Al}{2} as compared to the observations are shown in Figure~\ref{fig:lines}. Calculated emission line fluxes from both the models and observations are given in Table \ref{tab:lineflux}. The models reproduce \ion{H}{1} Ly$\alpha$, \ion{Si}{3}, \ion{O}{4}, the \ion{N}{5} doublet, and the \ion{C}{4} line at 1551 \AA \ within a factor of 2.5 of the observations. However, due to the simplified linear temperature structure, not all chromospheric lines are fit well. The models underpredict \ion{Si}{4}, \ion{O}{2}, and the \ion{C}{4} line at 1548 \AA \ by an order of magnitude. This implies that some EUV emission lines that form at similar temperatures in the lower transition region may also be underpredicted. The agreement between the synthetic spectra and each of the aforementioned lines can be improved by decreasing the temperature gradient in the transition region and reducing the thickness of the chromosphere. Making these adjustments, however, worsens the agreement with other FUV emission lines and elevates the EUV -- FUV pseudocontinuum by up to an order of magnitude. Since our final spectra are a good match to the observed UV continuum and many emission lines that form at temperatures found across the entire upper atmosphere, we conclude that the simplified temperature structures and predicted EUV spectra are good general approximations for each star.

The strongest emission line in the FUV spectrum is the Ly$\alpha$ resonance line. For most M dwarfs, it contributes up to 75\% of the total flux in the FUV region \citep{france2013}. Unfortunately, observations of the Ly$\alpha$ line for any star other than the Sun are heavily contaminated by both geocoronal airglow and interstellar hydrogen absorbing nearly all of the line core. In order to determine the intrinsic stellar line profile, observations must be reconstructed using interstellar parameters along the line of sight to the star \citep{wood2005,france2013,youngblood2016}. In the top panels of Figure \ref{fig:lines}, we show both the raw \textit{HST} observations and the \cite{youngblood2016} reconstructions. Our models closely reproduce the wings of Ly$\alpha$ in the raw \textit{HST} observation, but present with a self reversed core that results in calculated Ly$\alpha$ line fluxes 1.6 -- 2.5 $\times$ less than the reconstructed profiles. While the absorption in the raw \textit{HST} observations is due to contamination, the central reversal in our models is a direct result of non-LTE effects.

In the ultraviolet spectrum, non-LTE effects become important as the source function deviates from the Planck function in both the chromosphere and transition region. When allowing for departures from LTE, optically thick lines that form at various depths in a stellar atmosphere (e.g. Ly$\alpha$, \ion{Mg}{2}, \ion{Ca}{2}) can present in emission with a self-reversed core. Observations of the Sun in both active and quiescent states show self-reversals in Ly$\alpha$, \ion{Mg}{2} \textit{h} \& \textit{k}, and \ion{Ca}{2} H \& K lines \citep{linsky1970, fontenla1988, staath1995, tian2009}. In M dwarf stars, inverted cores are observed in high resolution measurements of \ion{Mg}{2} \textit{h} \& \textit{k} \citep{france2013} and some \ion{Ca}{2} H \& K lines \citep{rauscher2006}. 

While \ion{Mg}{2} \textit{h} \& \textit{k} form at slightly cooler temperatures than Ly$\alpha$, they have similar line profiles in the observed solar spectrum \citep{donnelly1994, lemaire1998}. Due to their similarities in the Sun, observations of the \ion{Mg}{2} doublet have been used to estimate the shape of the central portion of the Ly$\alpha$ profile for M stars after fitting the ISM absorption \citep{wood2005}. Low instrument resolution, however, can mask the existence of an inverted core in \ion{Mg}{2} lines. For example, in \cite{wood2005}, \textit{HST} observations of \ion{Mg}{2} \textit{h} \& \textit{k} and the corresponding best fit Ly$\alpha$ profiles for five M stars do not include a central reversal, but when utilizing a high resolution grating, observations of GJ 832 from \citealt{france2013} do show a central reversal in \ion{Mg}{2} (Figure \ref{fig:PRD}). 

Contamination of the stellar Ly$\alpha$ emission line from the geocoronal feature can be circumvented if observing a high radial velocity target. For example, the M1 star, Kapteyn's Star, has a radial velocity (V = +245~km~s$^{-1}$) such that Ly$\alpha$ is Doppler shifted 0.99 \AA \ away from the geocoronal contribution. \cite{guinan2016} obtained high resolution \textit{HST} observations of the Ly$\alpha$ region of Kapteyn's Star and determined the stellar emission profile presented with a faint self-reversal in the line core. Analysis of the same spectrum by \cite{youngblood2016}, however, fit the line with a Gaussian-shaped profile with no self-reversal. Additional high resolution observations of Ly$\alpha$ from high radial velocity M stars (e.g. A. Schneider et al., in prep.) as well as high resolution \ion{Mg}{2} observations would help determine the actual Ly$\alpha$ line profile and intrinsic flux from M stars. 

Modeling efforts by \cite{fontenla2016} also predict inverted cores in both the \ion{Mg}{2} doublet and Ly$\alpha$ for GJ 832. The smaller depths of the central reversal in the Kapteyn's Star observation and the \cite{fontenla2016} model lead us to believe our models could be underpredicting the flux in the core of Ly$\alpha$ by up to 40\% if the intrinsic profile does not contain any central reversal. We attribute uncertainty in the model Ly$\alpha$ line profiles to the parameterization in the transition region and the lack of ambipolar diffusion and a corona in our model. \cite{peacock2019} found that this reversal is very sensitive to $\nabla T_{TR}$, with less steep temperature gradients resulting in shallower reversals in the Ly$\alpha$ profile. As mentioned in Section \ref{subsec:vis}, the base of the transition region is determined by where hydrogen becomes fully ionized. Adding a corona above the transition region leads to back irradiation from the 10$^6$ K plasma, photoionizing the lower layers through ambipolar diffusion and affecting the collisional rates where the core is forming.

\subsection{Near UV and Far UV Photometry}

Existing \textit{GALEX} FUV (1340--1811 \AA) and NUV (1687--3008 \AA) detections of the stars measured using the elliptical Kron aperture are presented in Table \ref{tab:uvfluxes}. We compute synthetic photometry for the models over the same wavelength ranges as the \textit{GALEX} FUV and NUV filter profiles and find that the calculated values match the FUV detections, but exceed the NUV due to the overestimated \ion{Fe}{2} lines from 2375~--~2630~\AA. We also note that M stars are UV active and the \textit{GALEX} and \textit{HST} observations were not taken contemporaneously. It is therefore not necessarily expected that the model that best reproduces the \textit{HST} data also match the \textit{GALEX} photometry within uncertainty.

\begin{figure}[th!]
    \centering
    \includegraphics[width=1.0\linewidth]{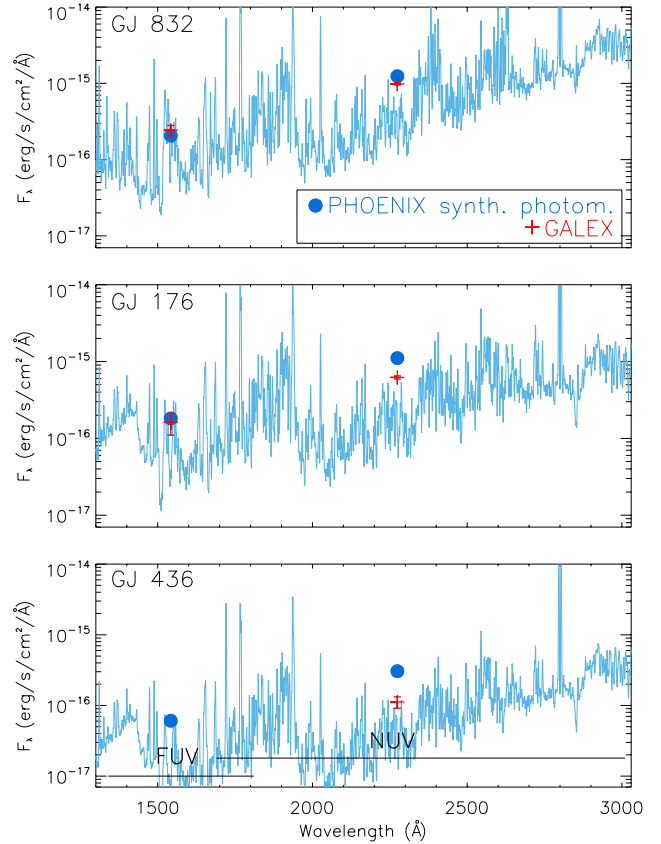}
    \caption{PHOENIX FUV--NUV spectra (blue) compared to \textit{GALEX} photometry (red, crosses), with calculated synthetic photometry over the same wavelengths as the \textit{GALEX} filter profiles plotted as blue circles. These wavelength ranges are indicated in the bottom panel. Values for the \textit{GALEX} detections are listed in Table \ref{tab:uvfluxes}.}
    \label{fig:galexphot}
\end{figure}

Model F$_{FUV}$ and F$_{NUV}$ for GJ 832 are 2.06 $\times$ 10$^{-16}$ erg cm$^{-2}$ s$^{-1}$ \AA \ and 12.41 $\times$ 10$^{-16}$ erg cm$^{-2}$ s$^{-1}$ \AA, respectively, falling within the uncertainty of the FUV detection but overpredicting the F$_{NUV}$ by a factor of 1.3. For GJ 176, F$_{FUV}$ = 1.81 $\times$ 10$^{-16}$ erg cm$^{-2}$ s$^{-1}$ \AA \ and F$_{NUV}$ = 11.09 $\times$ 10$^{-16}$ erg cm$^{-2}$ s$^{-1}$ \AA, matching the measured F$_{FUV}$ and overpredicting the F$_{NUV}$ detection by a factor of 1.7. Our model for GJ 436 overpredicts the singular F$_{NUV}$ detection by a factor of 2.6, F$_{NUV}$ = 3.05 $\times$ 10$^{-16}$ erg cm$^{-2}$ s$^{-1}$ \AA \ (Table \ref{tab:uvfluxes}). We plot this information in Figure~\ref{fig:galexphot}. 

The three prescribed parameters designating the temperature structure in the chromosphere and transition region for the three stars are nearly the same (Table \ref{tab:modelparam}). As a result of this degeneracy, the UV continuum slope in the models are very similar: F$_{NUV, G}$/F$_{FUV, G}$ = 5 -- 6. Calculating this ratio with the \textit{GALEX} detections for GJ~832 and GJ~176 both yield a ratio of $\sim$4,  supporting the plausibility that the upper atmospheres for these stars are likely analogous. 

\section{Extreme Ultraviolet Spectrum}\label{sec:euv}
Many emission features in the EUV (100 -- 1170 \AA) spectrum form in the upper chromosphere and transition region (e.g. \ion{He}{1} (584 \AA, 10$^{4.65}$ K) \ion{O}{5} (629.7~\AA, 10$^{5.3}$~K), and \ion{H}{1}~Ly$\beta$ (1025.7~\AA, 10$^{4.5}$~K) \citep{sim2005}), with highly ionized lines forming in the corona (e.g. \ion{Fe}{9} (171 \AA, 10$^6$ K) \citep{delzanna2014}). Continuum emission forms in the chromosphere and includes contributions from the \ion{H}{1} Lyman, \ion{He}{1} and \ion{He}{2} continua. 

\begin{deluxetable}{l|cc|ccc}[t]
\tablecaption{Band integrated UV flux densities in \\
(10$^{-16}$ erg cm$^{-2}$ s$^{-1}$ \AA$^{-1}$)}\label{tab:uvfluxes}
\tablehead{& 
\multicolumn2c{\textbf{GALEX Detections}} &
\multicolumn3c{\textbf{Models}} \\
& \colhead{F$_{FUV}$} & \colhead{F$_{NUV}$} & \colhead{F$_{EUV}$} & \colhead{F$_{FUV}$} & \colhead{F$_{NUV}$}}
\startdata
    GJ 832  & 2.43 $\pm$ 0.37 & 9.25 $\pm$ 0.31 & 14.74 & 2.06 & 12.41 \\ 
     GJ 176 & 1.62 $\pm$ 0.52 & 6.22 $\pm$ 0.30 & 17.04 & 1.81 & 11.09  \\
     GJ 436  & $\cdots$ & 1.12 $\pm$ 0.20 & 6.06 & 0.60 & 3.05 
\enddata
\tablecomments{\textit{GALEX} FUV and NUV photometric detections measured using the Kron aperture. Synthetic photometry from the PHOENIX models is calculated over $\lambda_{EUV}$ = 100--1170 \AA \ and the same wavelengths as the \textit{GALEX} FUV ($\lambda_{FUV}$ = 1340--1811 \AA) and NUV ($\lambda_{NUV}$ = 1687--3008 \AA) filter profiles.}
\end{deluxetable}

\begin{figure*}[th!]
    \centering
    \includegraphics[scale=1.1]{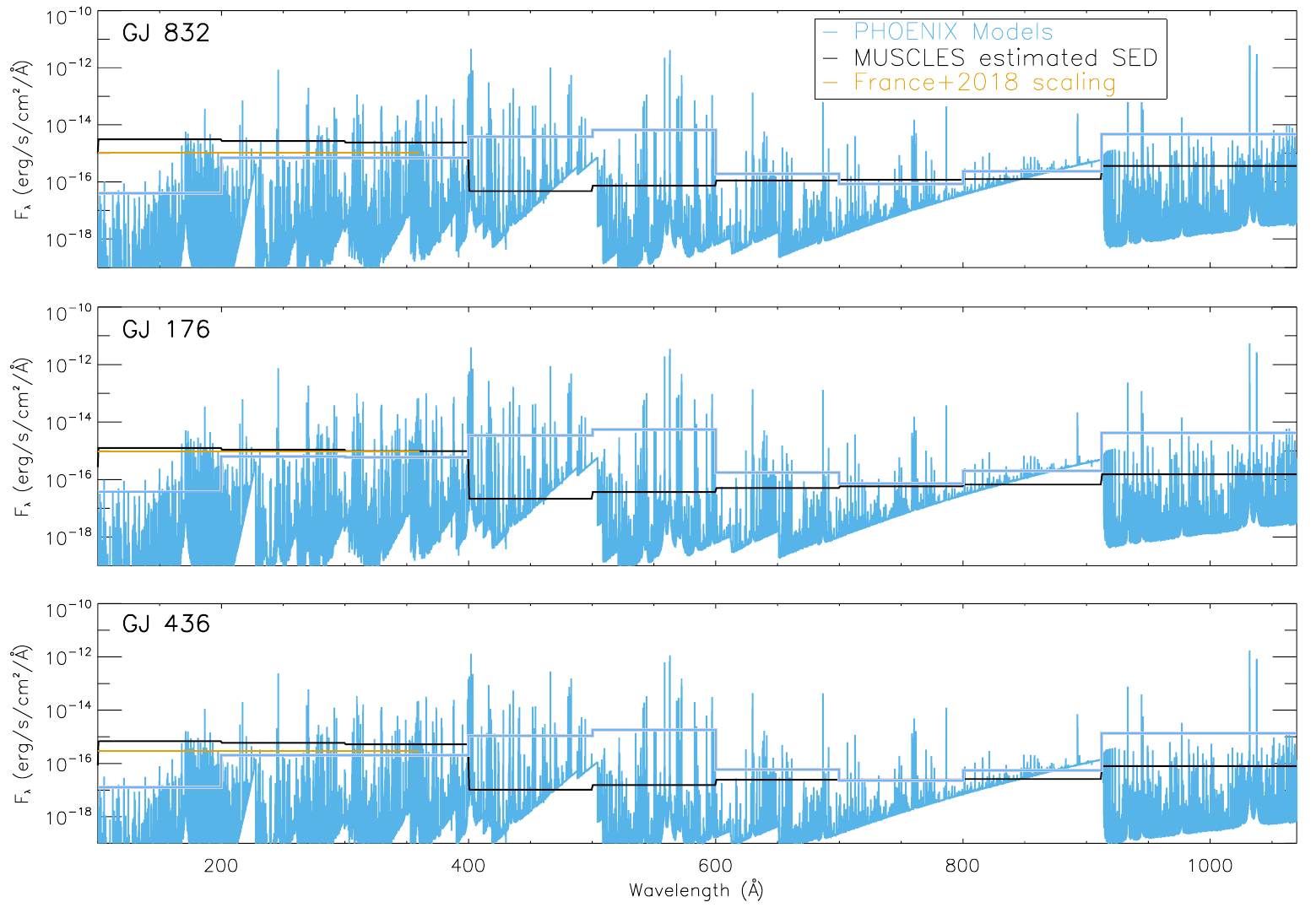}
    \caption{Full resolution EUV synthetic spectra and corresponding average flux densities in 100 \AA \ wavelength bands (blue). Estimated EUV flux densities in 100 \AA \ wavelength bands calculated using $F_{EUV}/F_{Ly\alpha}$ scaling relationships from \cite{linsky2014} in the MUSCLES SEDs are overplotted in black. F$_{EUV}$ estimates from \cite{france2018} are overplotted in orange.}
    \label{fig:EUVspec}
\end{figure*}

Our synthetic EUV spectra are presented in full resolution ($\Delta\lambda <$ 0.1 \AA) in Figure \ref{fig:EUVspec} with noticeable bound-free edges of \ion{H}{1} (912~\AA), \ion{He}{1} (504~\AA), \ion{N}{2} (418~\AA), \ion{K}{2} (392~\AA), \ion{O}{2} (353~\AA), \ion{He}{2} (228~\AA). In addition to the many non-LTE emission features, also prominent are narrow, but very bright emission features for species not included in the non-LTE set: \ion{Fe}{7} (246 \AA), \ion{Ne}{4} (402 \AA), \ion{Ca}{5} (558 \AA) and \ion{O}{6} (1031, 1038 \AA). The strength of the lines computed in the LTE, particularly those near 400 \AA \ and 550 \AA \ and the \ion{O}{6} lines near 1030 \AA \ are likely overpredicted by up to a factor of ten. On account of these lines, we suggest that the fluxes in the 400 -- 600 \AA \ and 1000 -- 1050 \AA \ regions be taken as upper limits. 

The current models do not include a corona, and therefore underpredict the flux from lines that form at temperatures greater than 2 $\times$ 10$^5$ K. While the continuum and many EUV emission lines found between 200 -- 1170 \AA \ form below this temperature, the 100 -- 200 \AA \ wavelength range in the \textit{EUVE} spectrum of the M2 flare star, AU Mic, is filled with Fe {\small\rmfamily XIX -- XXIII\relax} and Cr {\small\rmfamily XVIII -- XXI\relax} lines which form at coronal temperatures \citep{fossi1996}. 

As a test to quantify the amount of EUV flux our synthetic spectra could be underestimating, we compared our models to versions with example coronal spectra added to them. We computed coronal spectra of AU Mic (high activity) and the quiet Sun (low activity) with CHIANTI version 9.0 \citep{dere2019}, using the Differential Emission Measures (DEMs) of both objects from the CHIANTI database \citep{delzanna2002}. We truncated the temperature minimums at 2 $\times$ 10$^5$ K so that there was no duplication of the PHOENIX model structures. We scaled the computed AU Mic spectrum and the quiet Sun spectrum to GJ 832, GJ 436, and GJ 176, and added the CHIANTI spectra to each PHOENIX spectrum. AU Mic is an active young M star (12 Myr) with elevated levels of FUV and NUV emission \citep{Robinson2001}, and likely emits more EUV flux than GJ 832, GJ 436, and GJ 176. The addition of a scaled AU Mic coronal spectrum to our models increases the EUV flux in the 100 -- 200 \AA \ region by a factor of 60, but only slightly impacts on the rest of the EUV spectrum, increasing the integrated flux over 100 \AA \ wavelength bands by factors of 1 -- 5. Adding the scaled coronal spectrum of the quiet Sun to our models changes the EUV flux in each wavelength band by less than a factor of two, except in the 100 -- 200 \AA \ band, which increases by a factor of five. Integrating over 100 -- 1170 \AA, the coronal flux contribution from either DEM increases the overall EUV flux by 4 -- 45\%. Without the flux contribution from coronal lines, our models underpredict the EUV spectrum at $<$~200~\AA \ by less than two orders of magnitude and by less than a factor of five at wavelengths $>$~200~\AA. In a future paper, we will incorporate both a corona and the associated ambipolar diffusion to our models, exploring temperature structures tailored for the field age M stars.

\section{Discussion}\label{sec:discussion}

Since most of the EUV spectrum is unobservable due to interstellar contamination, there are no observations of the stars in this study to directly compare our synthetic spectra against. Here, we compare our models to existing EUV estimates for the target stars calculated with empirical scaling relationships or semiempirical stellar models and discuss the uncertainties associated with each method:

\subsection{Comparison of PHOENIX EUV Spectra to EUV Empirical Scaling Relationships}

In the MUSCLES spectral energy distributions (SEDs), \cite{youngblood2016} calculated the EUV spectrum for each star using Ly$\alpha$ reconstructions in an F$_{EUV}$/F$_{Ly\alpha}$ scaling relationship from \cite{linsky2014}. This method predicts the EUV flux from 100 -- 1070 \AA \ in 100 \AA \ wavelength bands and is derived from a combination of observations and models. The scalings for wavelengths $<$400 \AA \ are calculated from \textit{EUVE} observations of M stars, while those from 912 -- 1170 \AA \ are determined from \textit{FUSE} observations of K5 -- M5 stars. The relationships for wavelengths between 400 -- 912 \AA \ are computed from the \cite{fontenla2013} solar models and may not be appropriate for M stars. To compute the EUV flux using this scaling relationship, observations of the Ly$\alpha$ line must be reconstructed, contributing additional uncertainty to the predicted values. The F$_{EUV}$/F$_{Ly\alpha}$ estimated SEDs are plotted in black in Figure \ref{fig:EUVspec} and have an estimated accuracy of $\sim$20\% \citep{youngblood2016}. For ease of comparison, we have overplotted the average flux densities for our models in 100 \AA \ wavelength bands in blue. We find that this method is generally consistent with the synthetic spectra from 200 -- 400 \AA \ and 600 -- 900 \AA \ with fluxes agreeing within 40\% in these regions.

\cite{france2018} used \textit{EUVE} and \textit{HST} spectra of 104 F--M stars to develop a scaling relationship for F(90~--~360~\AA) based on \ion{N}{5} or \ion{Si}{4} observations. While the FUV emission lines used in this relationship do not rely on reconstructions, uncertainties in correcting for the ISM absorption in the \textit{EUVE} spectra and difficulties in measuring the continuum flux at these wavelengths result in uncertainties in the predicted EUV flux. The EUV fluxes for the target stars calculated with this scaling relationship are plotted in orange in Figure \ref{fig:EUVspec} and are estimated to be accurate within a factor of two. Comparing the \cite{france2018} scaling to the models across the full 90 -- 360 \AA \ range, the fluxes agree within factors of 0.95 -- 2.2. 

We also compare our model for GJ 436 to an EUV flux calculated in an F$_{EUV}$/F$_{X}$ scaling relationship derived in \cite{chadney2015}. This scaling is derived from averaged solar EUV and X-ray observations taken daily over an 11 year period and estimates an EUV flux for 124 -- 912 \AA. \cite{ehrenreich2015} calculated F$_{EUV}$~=~2.34 $\times$ 10$^{-13}$ erg cm$^{-2}$ s$^{-1}$ for GJ 436 using X-ray observations of the star. The \cite{linsky2014} Ly$\alpha$ scaling relationship is also derived from solar EUV observations, and yields a similar F(124 -- 912 \AA) = 2.1 $\times$ 10$^{-13}$ erg cm$^{-2}$ s$^{-1}$ \citep{youngblood2016}. Calculating the flux across the same wavelengths for our synthetic spectrum for GJ 436 yields a twice larger value of F$_{EUV}$~=~4.34~$\times$~10$^{-13}$~erg cm$^{-2}$~s$^{-1}$.

Empirical scaling relationships allow for general estimates of the unobservable EUV flux using observations from accessible wavelengths, but are derived using either observations of the Sun or from a wide range of spectral types. They are hindered by uncertainties in measuring the ISM absorption as well as noncontemporaneous observations at wavelengths known to display significant variability. They are further limited in both wavelength resolution and versatility. Conversely, semiempirical stellar models can predict realistic EUV spectra at the high resolution needed for detailed studies of the photochemistry and escape in exoplanet atmospheres and have the versatility to be used for a range of spectral type and activity state.

\subsection{Comparison of PHOENIX and Solar-Stellar Radiation Physical Modeling tools GJ 832 Models}

\cite{fontenla2016} adapted their 1D non-LTE semiempirical solar atmosphere model (SRPM) \citep{fontenla2015} that reproduces observations of the Sun, including UV spectra, to compute the stellar spectrum of GJ 832 using their Solar-Stellar Radiation Physical Modeling (SSRPM) tools. Similar to PHOENIX, their models are comprised of a modified thermal structure in the upper atmosphere added to an initial photospheric model with similar luminosity and spectral type as GJ 832. The temperature-pressure profile is based on their SRPM model and is divided into a chromosphere and lower transition region computed in a plane-parallel approximation and an upper transition region and coronal model computed in a spherically-symmetric approximation. The coronal SSRPM model extends to hotter temperatures than our PHOENIX model (0.2 MK), to their maximum coronal temperature peaking at 2.7 MK.

In both the PHOENIX and SSPRM models, solar elemental abundances are assumed, and the calculations include several species in full non-LTE radiative transfer in addition to millions of effectively thin background atomic and molecular lines. In our PHOENIX model, we consider 73 atoms and ions in full non-LTE, compared to 55 species (including \ce{H-} and \ce{H2}) in the \cite{fontenla2016} SSRPM model.

\begin{figure}[t!]
    \centering
    \includegraphics[width=1.0\linewidth]{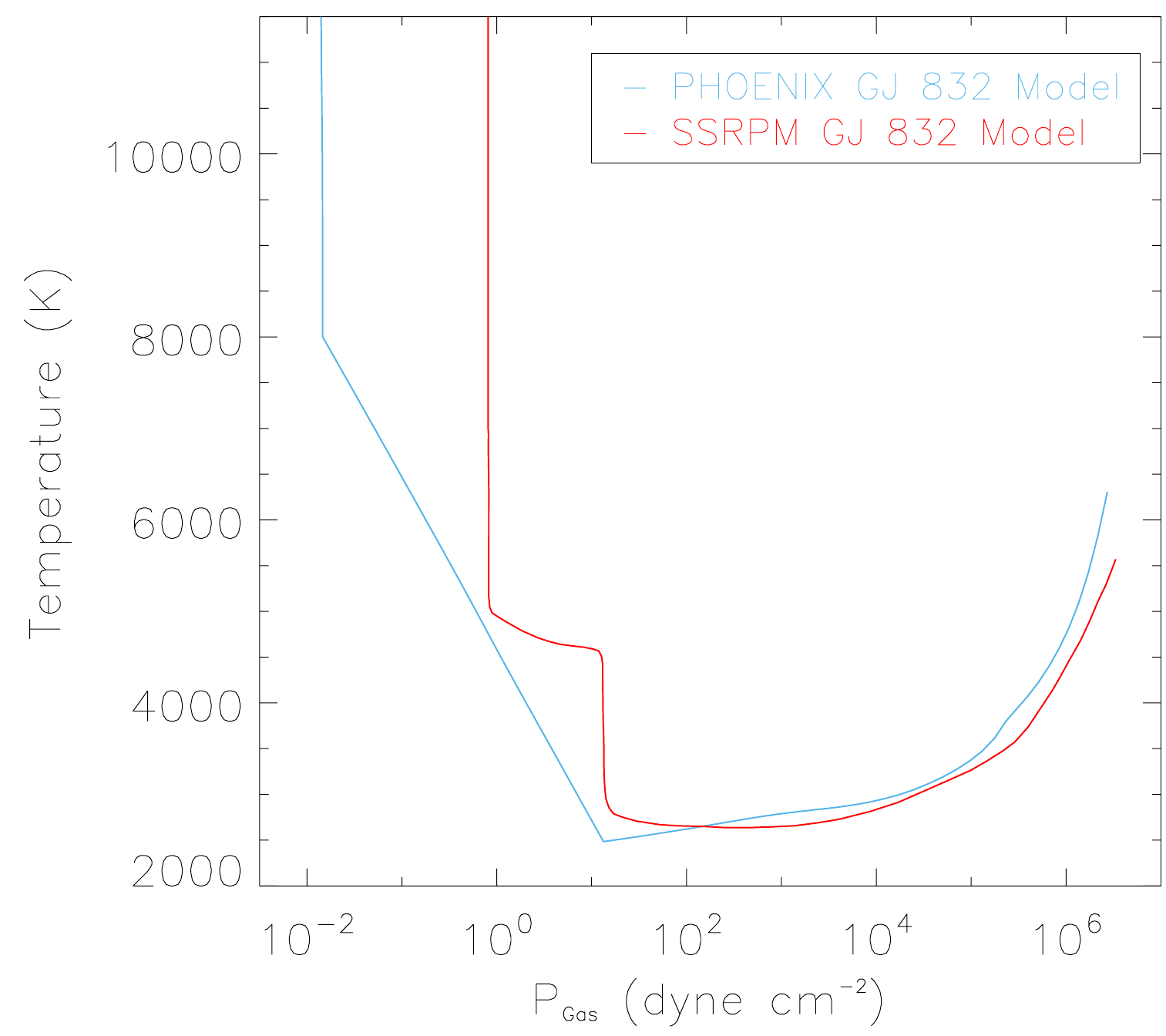}
    \caption{Temperature vs. gas pressure distributions in the lower transition region, chromosphere, and photosphere for the PHOENIX (blue) and the \cite{fontenla2016} SSRPM (red) GJ 832 models. The upper transition region in the PHOENIX model extends to 2 $\times$ 10$^5$ K. The final SSRPM GJ 832 model structure is coadded with a coronal temperature distribution that peaks around 2.7 $\times$ 10$^6$ K.}
    \label{fig:fontenlastructure}
\end{figure}

We compare the thermal structure of our GJ 832 model to the lower transition region, chromosphere and photosphere profile for the SSRPM GJ 832 model in Figure \ref{fig:fontenlastructure}. The GJ 832 models are both qualitatively and quantitatively very similar in the photosphere, but the chromospheric structure and onset of the thermally unstable transition region differ. The thermal structure in the SSRPM model has a steep temperature rise in the lower chromosphere near P$_{gas}$ $\approx$ 15 dynes cm$^{-2}$ followed by a near-constant temperature plateau in the upper chromosphere similar to their solar model. The temperature plateau results from the balance of radiative losses with non-radiative heating, and is where singly ionized metals are the dominant stages of ionization \citep{linsky2017}. Our model employs a linear temperature rise in log(column mass) for all chromospheric layers, which corresponds to a linear rise in log(P$_{gas}$) for this model, and similarly begins near P$_{gas}$ $\approx$ 15 dyne cm$^{-2}$.

\begin{figure}[t!]
    \centering
    \includegraphics[width=1.0\linewidth]{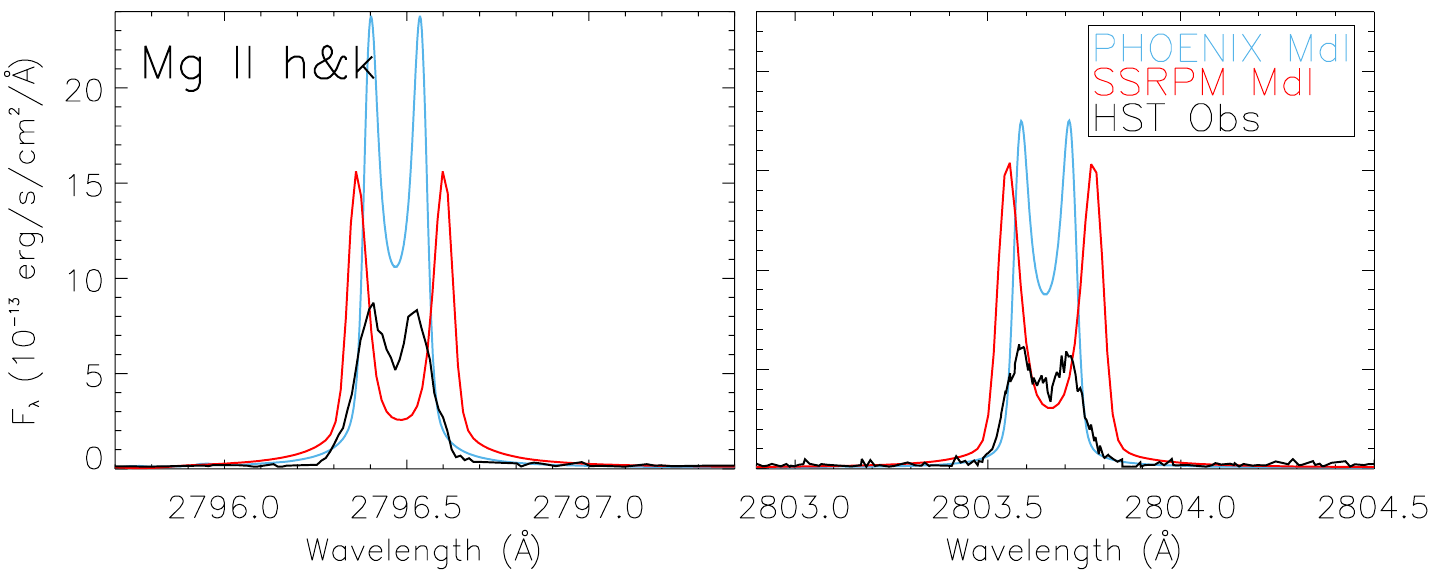}
    \caption{\ion{Mg}{2} \textit{h} \& \textit{k} profiles from the PHOENIX (blue) and SSRPM (red) models of GJ 832 compared to the high resolution STIS E230H observation (black). The observations have not been corrected for interstellar absorption.}
    \label{fig:fontenlamg}
\end{figure}

As described in Section \ref{sec:compobs}, the \ion{Mg}{2} doublet is an important diagnostic for the thermal profile in the chromosphere. We compare our PHOENIX model \ion{Mg}{2} \textit{h} \& \textit{k} profiles to the SSRPM model profiles, and the high resolution STIS E230H observations from the MUSCLES SED in Figure \ref{fig:fontenlamg}. Both models display central reversals due to non-LTE effects. Since the observed profiles have not been corrected for interstellar absorption a direct comparison to the core of the lines cannot be confidently made. The PHOENIX model computed in PRD more closely reproduces the observed line width than the SSRPM model. The wings of \ion{Mg}{2} begin forming around 3,000~K, indicating that the lower chromosphere may have a shallower slope than predicted in the SSRPM model.

\begin{figure*}[t]
    \centering
    \includegraphics[]{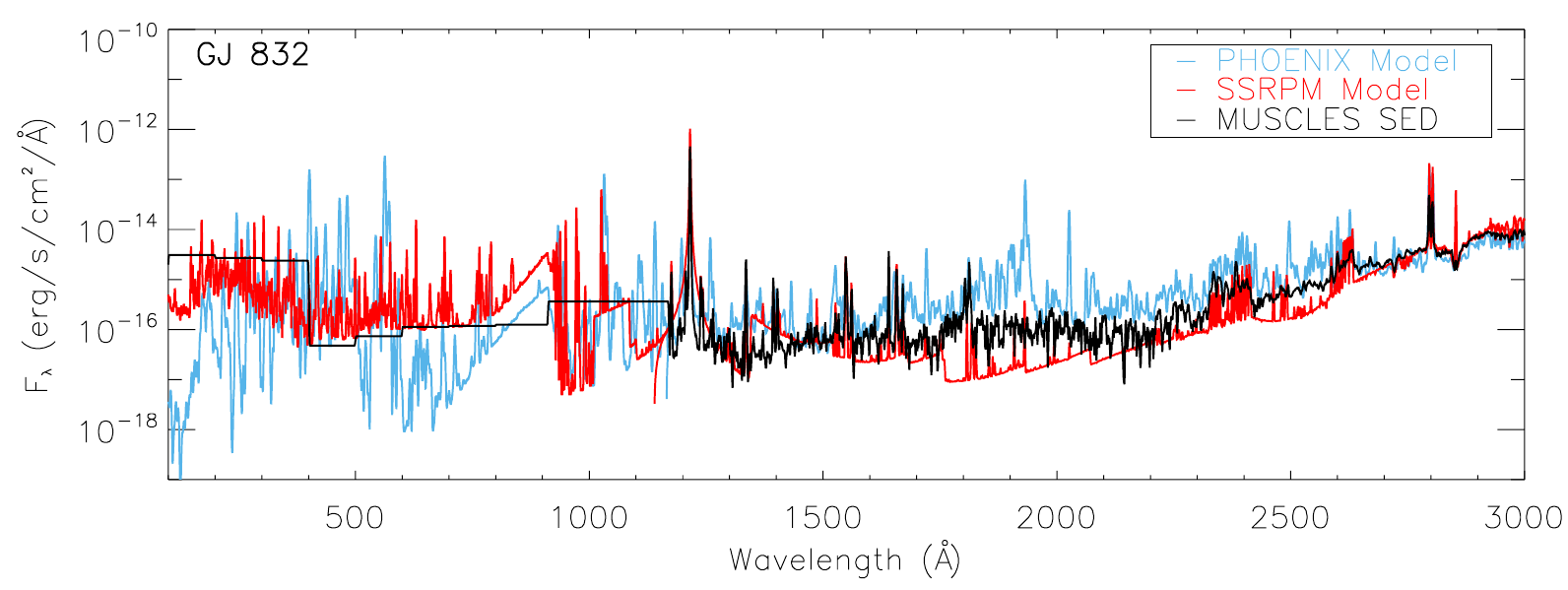}
    \caption{Comparison of EUV--NUV spectra for GJ 832. The PHOENIX model spectrum (blue) and \cite{fontenla2016} SSRPM model spectrum (red) have been degraded to the same 1 \AA \ resolution as the MUSCLES SED (black).}
    \label{fig:fontenlacomp}
\end{figure*}

Above the chromosphere, both models simulate a transition region with a steep temperature gradient governed by matching emission lines in the observed FUV spectrum. In our model, we set the temperature at the base of the transition region to 8,000~K, based on when the dominant cooling agent, neutral hydrogen, becomes fully ionized. In the SSRPM model, the transition region begins at 5,000~K. The likely reason for this discrepancy is the inclusion of a corona and ambipolar diffusion in the SSRPM tools, which are not yet included in PHOENIX. Using their solar model, \cite{fontenla1990} analyzed the energy balance of radiative losses with the downward flow of conductive heat and hydrogen ionization energy due to ambipolar diffusion from the corona. The authors found that the radiative losses were mainly due to hydrogen, and that ambipolar diffusion is greatly important in determining the hydrogen ionization in the lower transition region.

Figure \ref{fig:fontenlacomp} shows the PHOENIX and SSRPM synthetic EUV -- NUV spectra compared to the MUSCLES SED for GJ 832. We find good general agreement between all spectra long-ward of 200 \AA.  The SSRPM EUV spectrum from 100 -- 912 \AA \ is comprised of 36.4\% emission from the chromosphere and lower transition region model, and 63.3\% emission from the upper transition region and corona model. As compared to the SSRPM model, we estimate that the lack of coronal flux in our PHOENIX model may contribute up to an 80\% underprediction in only the 100 -- 200 \AA \ range. While the maximum temperature in our model does not extend to the same $\sim$ 10$^6$ K coronal temperatures, we find that the EUV spectrum from 200 -- 1170 \AA \ compares well to the SSRPM model, with F($\lambda$200--1170 \AA) =  1 $\times$ 10$^{-12}$ erg cm$^{-2}$ s$^{-1}$ for our model compared to F($\lambda$200--1170\AA) =  4 $\times$ 10$^{-13}$ erg cm$^{-2}$ s$^{-1}$ for that from \cite{fontenla2016}. 

Integrating over 100 -- 912 \AA, we find close agreement between our computed EUV luminosity: logL$_{EUV}$ = 27.31 erg s$^{-1}$, with the SSRPM model: logL$_{EUV}$ = 27.26 erg s$^{-1}$ and two other methods described in \cite{fontenla2016}. \cite{sanzforcada2011} calculated a predicted EUV luminosity for GJ 832 using \textit{ROSAT} X-ray observations. These observations give an L$_{X}$ 3.3 times higher than \textit{XMM-Newton} observations, indicating increased activity during the \textit{ROSAT} measurement. \cite{fontenla2016} scaled the predicted luminosity from \cite{sanzforcada2011} by this factor of 3.3 to yield logL$_{EUV}$ = 27.38 erg s$^{-1}$. Finally, the calculated EUV luminosity from the F$_{EUV}$/F$_{Ly\alpha}$ derived spectrum in the MUSCLES SED is logL$_{EUV}$ = 27.39 erg s$^{-1}$. The similarities of the predicted EUV luminosities for this star using a variety of methods and observations gives us confidence that all four techniques can be used to estimate the broadband EUV flux, however, the semiempirical models will provide the fine spectral resolution necessary for detailed studies of exoplanet atmospheres.

\section{Conclusions}\label{sec:conclusions}

We present high resolution EUV -- IR synthetic spectra of three early M planet hosts: GJ 832, GJ 176, and GJ 436. These synthetic spectra reproduce UV and visible spectral observations and UV, visible, and near-IR photometric detections and predict EUV fluxes similar to the active Sun. The models do not include absorption from the interstellar medium and can therefore be directly applied to investigations of photochemistry and stability of exoplanet atmospheres.

The temperature profiles for the models consist of a linear structure in the chromosphere and transition region. We find that nearly the same set of parameters ($\nabla$T$_{TR}$ = 10$^9$ K dyne$^{-1}$ cm$^2$, m$_{TR}$ = 10$^{-6.5}$ g cm$^{-2}$, m$_{Tmin}\simeq$10$^{-3.5}$ g cm$^{-2}$) best reproduces the UV observations for all three stars, suggesting that early M type stars may have similar thermal structures in their upper atmospheres. These similarities, however, could also be a result of the simplified thermal structure averaging out small differences between the stars. Cool stars are highly active in their upper atmospheric layer with locally active regions of enhanced or depressed EUV flux. For example, spatially varying solar atmospheric features, for which there are high resolution spectra available, are modeled with several different thermal structures \citep{fontenla2011}. Our simplified models provide general approximations of the EUV spectrum.

We find that our simplified structure produces similar continuum and line fluxes (for wavelengths greater than 200 \AA) to the GJ 832 spectrum computed with the SSRPM semiempirical stellar atmosphere code, which includes a corona \citep{fontenla2016}. The thermal structures in each model are significantly different in the upper-most atmospheric layers, with the onset of the transition region beginning at a pressure nearly 100$\times$ lower and a temperature 3,000 K hotter in the PHOENIX model. To improve our general understanding of the upper atmospheric temperature-pressure profile and prediction of EUV fluxes, we will extend our thermal structures to coronal temperatures and quantify the importance of ambipolar diffusion in future work.

Starting in the early 2020s after \textit{HST} stops UV operations, there will be an observation gap for FUV and NUV spectroscopy, in addition to the current gap in EUV observations for any star other than the Sun. During this time, there will be no instrument available to follow up with UV observations of newly discovered planet host stars. The expansive database of archival \textit{GALEX} UV photometry for hundreds of M stars ranging in age and spectral type can be used to guide upper-atmosphere models such as these. These models predict realistic high resolution spectra across unobservable wavelengths and are important for furthering our understanding of the effects of high energy radiation on planets orbiting M stars.

\acknowledgements

This work was supported by NASA Headquarters under the NASA Earth and Space Science Fellowship Program-Grant NNX15AQ94H. E.S. acknowledges support from the NASA Habitable Worlds grant NNX16AB62G. We also gratefully acknowledge support from NASA HST Grant HST-GO-14784.001-A. An allocation of computer time from the UA Research Computing High Performance Computing (HPC) at the University of Arizona is gratefully acknowledged. A portion of the calculations presented here were performed at the Höchstleistungs Rechenzentrum Nord (HLRN), and at the National Energy Research Supercomputer Center (NERSC), which is supported by the Office of Science of the U.S. Department of Energy under Contract No. DE-AC03-76SF00098. We thank all these institutions for a generous allocation of computer time. E.B. acknowledges support from NASA Grant NNX17AG24G. This work has made use of data from the European Space Agency (ESA) mission {\it Gaia} (\url{https://www.cosmos.esa.int/gaia}), processed by the {\it Gaia} Data Processing and Analysis Consortium (DPAC, \url{https://www.cosmos.esa.int/web/gaia/dpac/consortium}). Funding for the DPAC has been provided by national institutions, in particular the institutions participating in the {\it Gaia} Multilateral Agreement. 

\software{PHOENIX \citep{hauschildt1993,hauschildt2006,baron2007}}

\end{document}